\documentclass[aps,prc,twocolumn,superscriptaddress,10pt,english,preprintnumbers,nofootinbib]{revtex4-1}
\pdfoutput=1
\usepackage[T1]{fontenc}
\usepackage[utf8]{inputenc}
\usepackage{epsfig}
\usepackage{graphicx}
\usepackage{xcolor}
\usepackage{latexsym}
\usepackage{textcomp}
\usepackage{amssymb}
\usepackage[colorlinks=true,linkcolor=blue,citecolor=blue, urlcolor=blue]{hyperref} 
\usepackage{amsfonts,amsthm,amstext,amscd}
\usepackage{amsmath}
\usepackage{mathtools}
\usepackage{xspace}
\usepackage{booktabs}
\usepackage{soul}

\newcommand{\PbPb}         {\mbox{Pb--Pb}}
\newcommand{\Bfivenn}       {$\sqrt{s_{\mathrm{NN}}}=5.02$\,TeV}
\newcommand{\Btwonn}       {$\sqrt{s_{\mathrm{NN}}}=2.76$\,TeV}

\newcommand{\GeVmass}      {GeV/$c^2$}

\begin{document}
\title{Effective temperatures of the QGP from thermal photon and dilepton production}
\author{Olaf Massen}
\affiliation{Institute for Gravitational and Subatomic Physics, Utrecht University, 3584 CC Utrecht, The Netherlands}
\affiliation{Nikhef, Science Park 105, 1098 XG Amsterdam, The Netherlands}
\author{Govert Nijs}
\affiliation{Theoretical Physics Department, CERN, CH-1211 Gen\`eve 23, Switzerland}
\author{Mike Sas}
\affiliation{Institute for Gravitational and Subatomic Physics, Utrecht University, 3584 CC Utrecht, The Netherlands}
\affiliation{Nikhef, Science Park 105, 1098 XG Amsterdam, The Netherlands}
\affiliation{European Organization for Nuclear Research (CERN), Geneva, Switzerland}
\author{Wilke van der Schee}
\affiliation{Theoretical Physics Department, CERN, CH-1211 Gen\`eve 23, Switzerland}
\affiliation{Institute for Theoretical Physics, Utrecht University, 3584 CC Utrecht, The Netherlands}
\affiliation{Nikhef, Science Park 105, 1098 XG Amsterdam, The Netherlands}
\author{Raimond Snellings}
\affiliation{Institute for Gravitational and Subatomic Physics, Utrecht University, 3584 CC Utrecht, The Netherlands}
\affiliation{Nikhef, Science Park 105, 1098 XG Amsterdam, The Netherlands}
\begin{abstract}
Thermal electromagnetic radiation is emitted by the quark-gluon plasma (QGP) throughout its space-time evolution, with production rates that depend characteristically on the temperature. We study this temperature using thermal photons and dileptons using the \emph{Trajectum} heavy ion code, which is constrained by Bayesian analysis. In addition we present the elliptic flow of both the thermal photons and thermal dileptons including systematic uncertainties corresponding to the model parameter uncertainty. We give a comprehensive overview of the resulting effective temperatures $T_{\rm eff}$, obtained from thermal photon transverse momentum and thermal dilepton invariant mass distributions, as well as the dependence of $T_{\rm eff}$ on various selection criteria of these probes. We conclude that the $T_{\rm eff}$ obtained from thermal photons is mostly insensitive to the temperature of the QGP with a value of $T_{\rm eff} \sim 250$--300\,MeV depending on their transverse momentum, almost independent of collision centrality. Thermal dileptons are much better probes of the QGP temperature as they do not suffer from a blue shift as their invariant mass is used, allowing for a more precise constraint of the QGP temperature during different stages of the evolution of the system. By applying  selection criteria on the dilepton transverse momentum and the invariant mass we are able to extract fluid temperatures on average times ranging from late emission ($\langle \tau \rangle = 5.6\,$fm$/c$) to very early emissions ($\langle \tau \rangle < 1.0\,$fm$/c$)\@. Furthermore, we show how these selection criteria can be used to map the elliptic flow of the system all throughout its evolution.
\end{abstract}

\preprint{CERN-TH-2024-213}

\maketitle

\section{Introduction}

The quark-gluon plasma (QGP), created in ultrarelativistic heavy-ion collisions, is a strongly coupled state of matter consisting of deconfined quarks and gluons, which behaves as a near-perfect liquid \cite{Heinz:2013th, Busza:2018rrf}\@. Calculations using lattice Quantum Chromodynamics (QCD) have shown that there is a crossover from hadronic matter to a QGP around a deconfinement temperature of approximately $T_{\rm co}\approx 155\,$MeV \cite{Aoki:2006br, Bazavov:2011nk, HotQCD:2014kol, Guenther:2020jwe}\@. Anisotropic flow measurements have had a crucial role in our understanding of the systems created in heavy-ion collisions. The azimuthal anisotropy of produced particles provides information on the spatial anisotropies present in the initial stages of the collisions and the subsequent hydrodynamic evolution with characteristically small specific shear and bulk viscosities \cite{Ollitrault:1997di, Bass:1998vz, Voloshin:2008dg, Bernhard:2019bmu}\@. There has been a long search to determine the QGP temperature directly from experiments \cite{Busza:2018rrf}\@. Within hydrodynamics this is challenging, since the equations of motion only rely on the pressure versus energy density and not on the temperature directly, but relatively successful estimates along this route are possible \cite{Gardim:2019xjs}\@. An even more direct way to estimate the temperature makes use of the thermal electromagnetic radiation, which is emitted throughout the QGPs lifetime and is the topic of this work \cite{Geurts:2022xmk, Shen:2016odt, Rapp:2016xzw}\@.

The QGP temperature $T$ is difficult to determine from hadrons because they are emitted around the deconfinement temperature and consequently do not contain much information on earlier times \cite{ALICE:2022wpn}\@. However, thermal electromagnetic radiation is able to escape unscathed, as their mean free path is much larger than the size of the system, and, as a result, carries information about the QGP temperature all throughout its evolution (see \cite{Peitzmann:2001mz, David:2019wpt, Geurts:2022xmk} for reviews)\@. This radiation is emitted by the plasma in the form of thermal photons and dileptons, the production rate of which scales as $T^4$ \cite{McLerran:1984ay, Arnold:2001ba, Arnold:2001ms, Arnold:2002ja}\@. 

Another way we can utilise this property of electromagnetic radiation is by studying the anisotropic flow of photons and dileptons at different average emission times. Anisotropic flow develops over time and, since electromagnetic radiation retains its information, it can provide direct information on the space-time evolution of the system and how the spatial anisotropies are transformed into momentum-space anisotropies.

For thermal photons, originating at leading order from quark-gluon Compton scattering and $q\bar{q}$-annihilation, the invariant yield as a function of transverse momentum follows a roughly exponential behaviour. This thermal photon yield can then be related to an effective temperature. Thermal dileptons originate from decays of virtual photons $(\gamma^{*}\rightarrow e^{+}e^{-})$ and are produced through $q\bar{q}$-annihilation within the plasma. For these an effective temperature can be obtained from the shape of the distribution of the virtuality of the photon, which is equal to the invariant mass of the dilepton pair. For both these probes, measurements at the Relativistic Heavy Ion Collider (RHIC) and the Large Hadron Collider (LHC) estimate the QGP effective temperature to be around 300\,MeV, with centre-of-mass energies ranging from 200\,GeV to 5.02\,TeV \cite{ALICE:2015xmh, STAR:2016use, PHENIX:2022rsx, PHENIX:2022qfp, ALICE:2023jef, STAR:2024bpc}\@. Without detailed theoretical calculations it remains difficult to interpret these effective temperatures \cite{Shen:2013vja, Gale:2012xq, Paquet:2022wgu, Paquet:2023bdx}\@.

Many theoretical studies have been done for both photon and dilepton production in relativistic heavy ion collisions \cite{Deng:2010pq, Vujanovic:2013jpa, Linnyk:2013hta, Gale:2014dfa, Bhattacharya:2015ada, Vujanovic:2016anq,  Endres:2016tkg, Hauksson:2017udm, Paquet:2017wji, Kasmaei:2018oag, Churchill:2023vpt, Churchill:2023zkk, Wu:2024pba}\@. The production rates have been determined at all stages of the collisions: prompt production \cite{Paquet:2015lta}, pre-equilibrium contributions \cite{Garcia-Montero:2023lrd}, the QGP contributions \cite{Arnold:2001ba, Arnold:2001ms} and the contributions from the hadron resonance gas \cite{Turbide:2003si, vanHees:2014ida, Linnyk:2015tha}\@.

Here we will present an implementation of the QGP emission of both photons and dileptons at LHC energies and determine the resulting spectra and anistropic flow. With this implementation we can then investigate the connection between the temperature of the QGP and its effective temperature as extracted from both thermal photon and dilepton production for \PbPb{} collisions at \Bfivenn{}\@. We study the full centrality dependence and different photon and dilepton selection criteria, and include systematic uncertainties based on the uncertainty of the model parameters obtained in a Bayesian analysis. In addition we present the anisotropic flow of these thermal probes and how this anisotropic flow develops over time. The results are obtained using the heavy-ion simulation code \emph{Trajectum}~\cite{Nijs:2020ors, Nijs:2020roc, trajectumcode}, which for this work was extended to include thermal photon and dilepton emission from the QGP\@.

\begin{figure}
    \centering
    \includegraphics[width=\columnwidth]{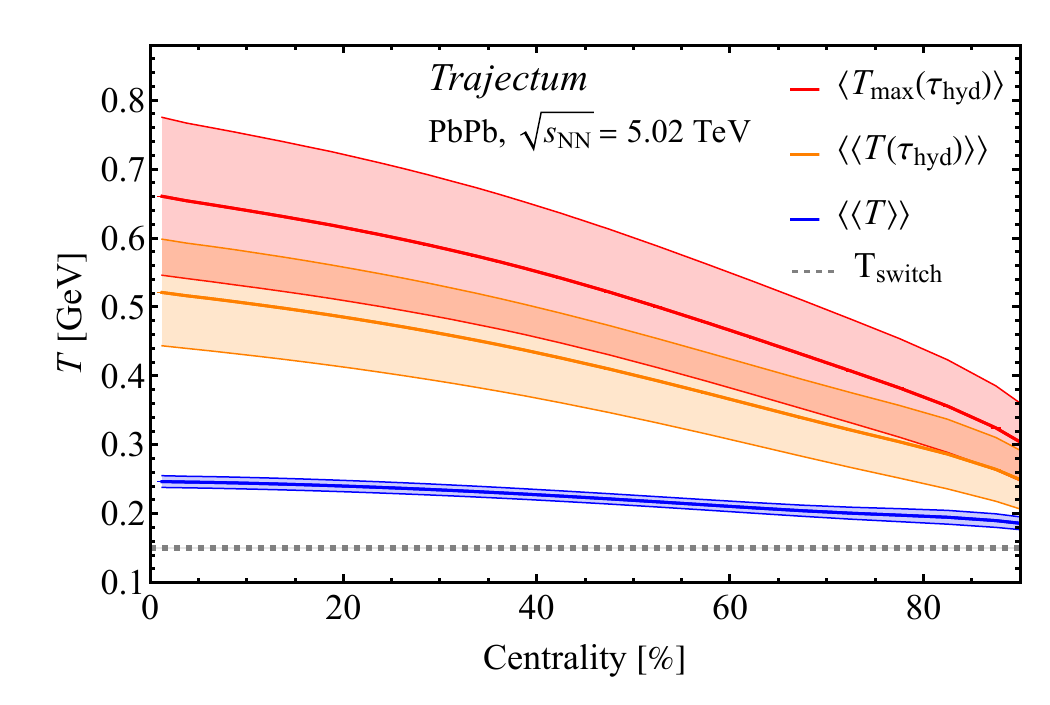}
    \caption{Averaged temperatures of the QGP as function of collision centrality along with the switching temperature $T_\text{switch}$\@ (see text for more details). The coloured bands correspond to the systematic uncertainty associated with the \emph{Trajectum} model.}
    \label{fig:temperaturesCentrality}
\end{figure}

\section{Temperatures in \emph{Trajectum}}

\emph{Trajectum} is a state-of-the-art hybrid model describing the evolution of a heavy-ion collision from the initial state up until the hadronic stage. The initial state is a generalised version of the T\raisebox{-.5ex}{R}ENTo~\cite{Moreland:2014oya} initial conditions, which is followed by a far-from-equilibrium stage that can interpolate between a weakly coupled and a strongly coupled description. After the pre-equilibrium stage, which ends at proper time $\tau_\text{hydro}$, the system is evolved according to second-order 2+1D boost invariant viscous hydrodynamics with temperature-dependent transport coefficients, until the system reaches a certain particlisation temperature ($T_\text{switch}$)\@. Then, the fluid is transformed into hadrons via the Cooper-Frye mechanism with viscous corrections according to the Pratt-Torrieri-Bernhard prescription \cite{Pratt:2010jt, Bernhard:2018hnz}\@. These hadrons can then be fed to a hadronic afterburner such as SMASH \cite{SMASH:2016zqf, SMASH}\@. The parameters of the \emph{Trajectum} model have been tuned to \PbPb{} collisions at $\sqrt{s_\text{NN}} = 2.76$ and 5.02\,TeV using Bayesian analysis \cite{Giacalone:2023cet}\@.

We determine the systematic uncertainty of the \emph{Trajectum} model by sampling twenty different parameter configurations, which are based on the posterior distribution obtained from the Bayesian analysis. For each parameter configuration we simulate new events to calculate all the relevant observables, and obtain the systematic uncertainty as the RMS of the points, where the statistical uncertainty is quadratically subtracted.

The temperature of the plasma is readily available within the \emph{Trajectum} framework, in contrast to experimental studies. To understand the temperature evolution throughout the system we studied the maximum temperature as well as the average temperature as weighted by the energy density both as a function of collision centrality and of proper time $\tau$\@. We take into account all the fluid cells with temperatures above the switching temperature $T_\text{switch}$, which is the reason why the temperatures are always strictly above $T_\text{switch}$\@.
In \emph{Trajectum} we determine the centrality interval by the number of charged particles produced at mid-rapidity, whereby the 0--1\% centrality interval corresponds to the highest multiplicity and smallest impact parameter. In ALICE this is typically determined at forward rapidity \cite{ALICE:2013hur,Nijs:2023bzv}, but for this study no significant differences are expected.

\begin{figure}
    \centering
    \includegraphics[width=\columnwidth]{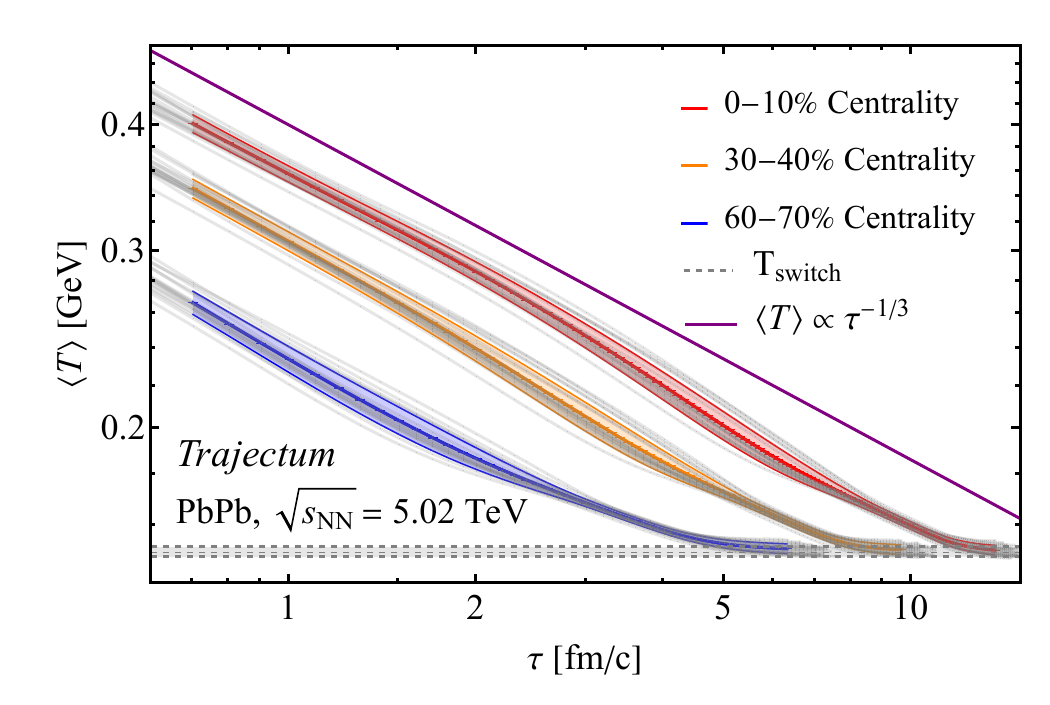}
    \caption{Averaged temperatures of the QGP as a function of time $\tau$, for three intervals in collision centrality corresponding to central, semi-central, and peripheral collisions. The coloured bands correspond to the systematic uncertainty associated with the \emph{Trajectum} model. In the plot we show the 20 parameter variations in gray, as well as a reference for $\langle T \rangle = \alpha \tau^{-1/3}$\@.}
    \label{fig:temperaturesTau}
\end{figure}
 
In Fig.~\ref{fig:temperaturesCentrality} we present the different temperatures $\langle T_\text{max}(\tau_{\text{hydro}})\rangle$, $\langle\langle T(\tau_{\text{hydro}})\rangle\rangle$, and $\langle \langle T \rangle \rangle$ as a function of collision centrality. $\langle T_\text{max}(\tau_{\text{hydro}})\rangle$ corresponds to the maximum temperature reached locally in the plasma at proper time $\tau_{\text{hydro}}$, averaged over all collisions. $\langle\langle T(\tau_{\text{hydro}})\rangle\rangle$ is the space-averaged temperature of the plasma at $\tau_{\text{hydro}}$, averaged over all collisions. Finally, $\langle \langle T \rangle \rangle$ is the space and time-averaged temperature of the plasma, again averaged over all collisions, which thus takes into account the full evolution of the system. All temperature averages are weighted using the energy density. As expected, the highest temperatures are observed in the most central collisions. The maximum temperature and its space averaged temperature at the start of the hydrodynamic evolution, $\langle T_\text{max}(\tau_{\text{hydro}})\rangle$ and $\langle\langle T(\tau_{\text{hydro}})\rangle\rangle$, show a strong centrality dependence. The space and time averaged temperature $\langle \langle T \rangle \rangle$ shows a milder centrality dependence and has less of a dependence on the choice of $\tau_{\text{hydro}}$, as the average is largely driven by the extended time at which the plasma is below $T \simeq 200$\,MeV\@.

It is important to note that the temperature of the plasma at $\tau_{\text{hydro}}$ depends strongly on the value of this particular parameter.\footnote{For a conformal free streaming plasma the energy density of the system scales as $1/\tau$, leading to a temperature dependence of $\tau^{-1/4}$\@. For an expansion described by ideal hydrodynamics the work term leads to a temperature evolution that is proportional to $\tau^{-1/3}$.} In \cite{Giacalone:2023cet}, $\tau_{\text{hydro}}$ is not well constrained at $\tau_{\text{hydro}} \approx 0.38 \pm 0.17\,$fm$/c$ and hence dominates the systematic uncertainties of both $\langle T_\text{max}(\tau_{\text{hydro}})\rangle$ and $\langle\langle T(\tau_{\text{hydro}})\rangle\rangle$\@.

Fig.~\ref{fig:temperaturesTau} shows $\langle T \rangle$ as a function of proper time $\tau$\@. Here, the systematic uncertainty on the temperature is typically less than 5\% and statistical uncertainties are shown but negligible. We note that the systematic uncertainty represents the uncertainty in the parameters, and not the variations in the averaging over collisions in a centrality class.
For more central collisions the nuclei have a larger overlapping area, which, combined with the increased density at the centre of the nuclei, leads to a larger and hotter plasma. Central collisions hence produce a QGP that is hotter and lives longer.

\section{Thermal photon and dilepton implementation}

For many years people have worked on the thermal production of electromagnetic probes \cite{McLerran:1984ay, Gale:1987ki, Strickland:1994rf, Chatterjee:2007xk}\@. Production rates of both photons and dileptons have been studied for a weakly-coupled plasma, using hard-thermal-loop effective field theory and kinetic theory, and a strongly coupled plasma, using holography \cite{Finazzo:2015xwa, Iatrakis:2016ugz}\@. In this work we study the first approach, implementing the next-to-leading order emission of thermal photons \cite{Ghiglieri:2013gia} and dileptons \cite{Ghiglieri:2014kma} for a weakly coupled QGP\@. These production rates have been embedded into the hydrodynamic framework of the \emph{Trajectum} model. For details on the theoretical calculation see \cite{Ghiglieri:2013gia, Ghiglieri:2014kma}\@. In thermal equilibrium the production rate of both the photons and dileptons is governed by the current-current correlation function $\Pi^{<}(K)$ of the electromagnetic current operator $J^{\mu}$:
\[
    \Pi^{<}(K) \equiv \int d^4X e^{-K\cdot X}\langle J^{\mu}(0)J_{\mu}(X)\rangle,
\]
where $K$ is the four-momentum of the photon, $X$ the spacetime coordinate and $\langle \cdot \rangle$ the expectation value in the QGP ensemble. The production rates per unit 4-volume $d\Gamma/d^4K$ as a function of the energy $k^0$ and momentum $\mathbf{k}$ in the local fluid rest frame are given by:
\begin{align*}
    \frac{dN_{\gamma}}{d^4Xd^3\bf{k}} \equiv \frac{d\Gamma_{\gamma}}{d^3\bf{k}} & = \left. \frac{1}{(2\pi)^3 2|\mathbf{k}|}\Pi^<(K)\right|_{k^0 = |\mathbf{k}|}, \\
    \frac{dN_{ll}}{d^4Xd^4K} \equiv \frac{d\Gamma_{ll}}{d^4K} & = \frac{2\alpha_\text{EM}}{3(2\pi)^4 M^2}\Pi^<(K)\\
    & \quad \times \mathcal{B}\left(\frac{m^2_l}{M^2} \right) \Theta((k^0)^2 - \mathbf{k}^2),
\end{align*}
for the thermal photons ($\gamma$) and thermal dileptons ($ll$) respectively, with $\alpha_{EM}$ the electromagnetic coupling constant and $ \Theta((k^0)^2 - \mathbf{k}^2)$ the Heaviside step function to ensure a positive invariant mass for the dileptons. In the case of the photons, the 4-momentum $K$ is taken to be on-shell, i.e.~$K^2 = 0$\@. In case of the dilepton production the 4-momentum $K$ is taken to be time-like with positive energy, such that $M^2=-K^2 > 4m_l^2$, with $M$ the mass of the virtual photon. In addition, the production is considered away from threshold where $M \gg m_l$, and therefore the phase-space factor $\mathcal{B}(x) = (1+2x)\sqrt{1-4x} \approx 1$, with $x=m^2_l/M^2$\@.

Within \emph{Trajectum}, thermal photons and dileptons are emitted purely during the hydrodynamic phase of the heavy-ion collision. The hydrodynamic evolution is performed on a grid of fluid cells. In each time step of the evolution we consider all fluid cells with a temperature $T>T_{\rm switch}$, where $T_{\rm switch}$ is the particlisation temperature of the plasma. The number of produced photons and dileptons per fluid cell is sampled from a Poisson distribution, with the average determined by the integrated production rates, the temperature in the fluid cell, and the 4-volume of the fluid cell in the local fluid rest frame. Each produced photon and dilepton is then assigned a random momentum $\bold{k}$, and energy $k^0$ for the dileptons, according to the differential production rates. Finally the momenta and energies are transformed from the local fluid rest frame to the lab frame.

In addition to the thermal photon production by the QGP, the thermal contribution of the hadron resonance gas to the photon production was taken into account, as calculated by the hadronic afterburner SMASH \cite{SMASH,SMASH:2016zqf}\@. In addition it is important to note that we only consider thermal photons with $p_T > 0.1$\,GeV$/c$ and dileptons with both $p_T > 0.2$\,GeV$/c$ and $m_{ll} > 0.2$\,GeV$/c^2$, due to large theoretical uncertainties in the rates below these values \cite{Ghiglieri:2013gia}\@. Since the dileptons are not simulated in SMASH it is relatively straightforward to improve the statistics by using an increased value for $\alpha_{\rm EM}$ and reducing the final spectra accordingly.

\begin{figure}
    \centering
    \includegraphics[width=\columnwidth]{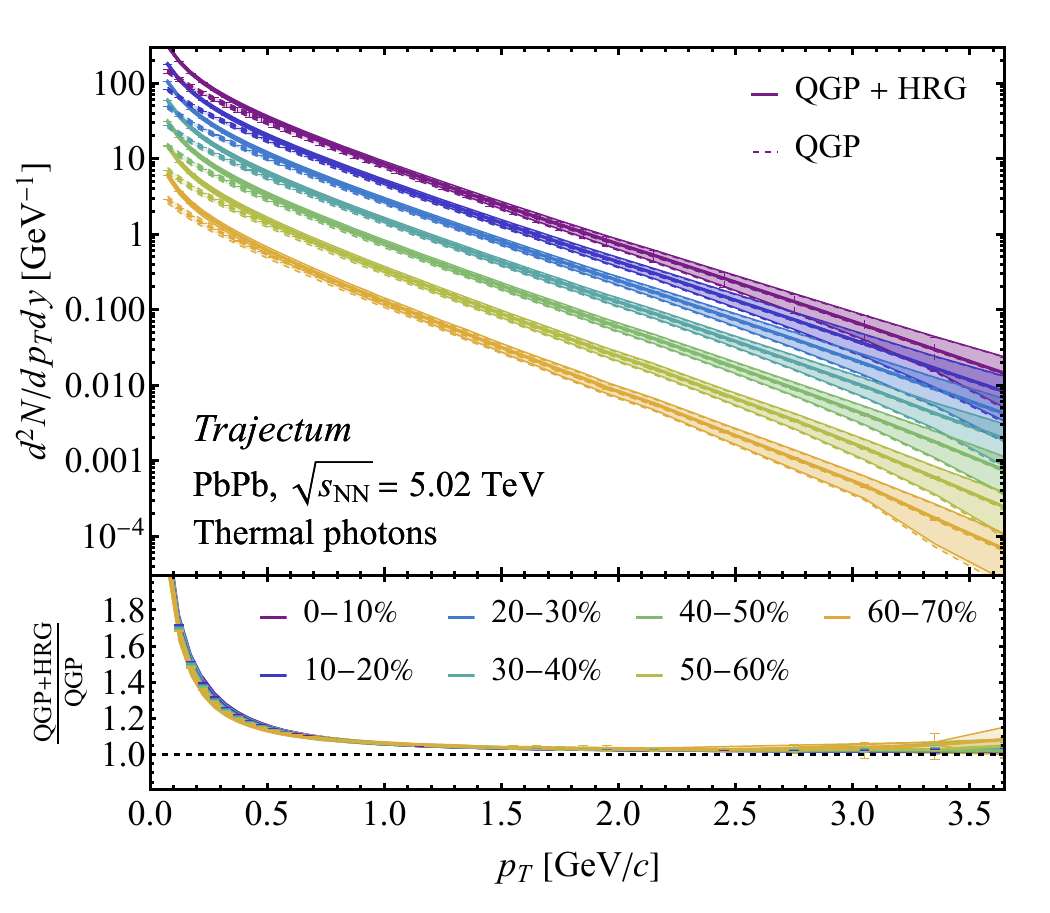}
    \includegraphics[width=\columnwidth]{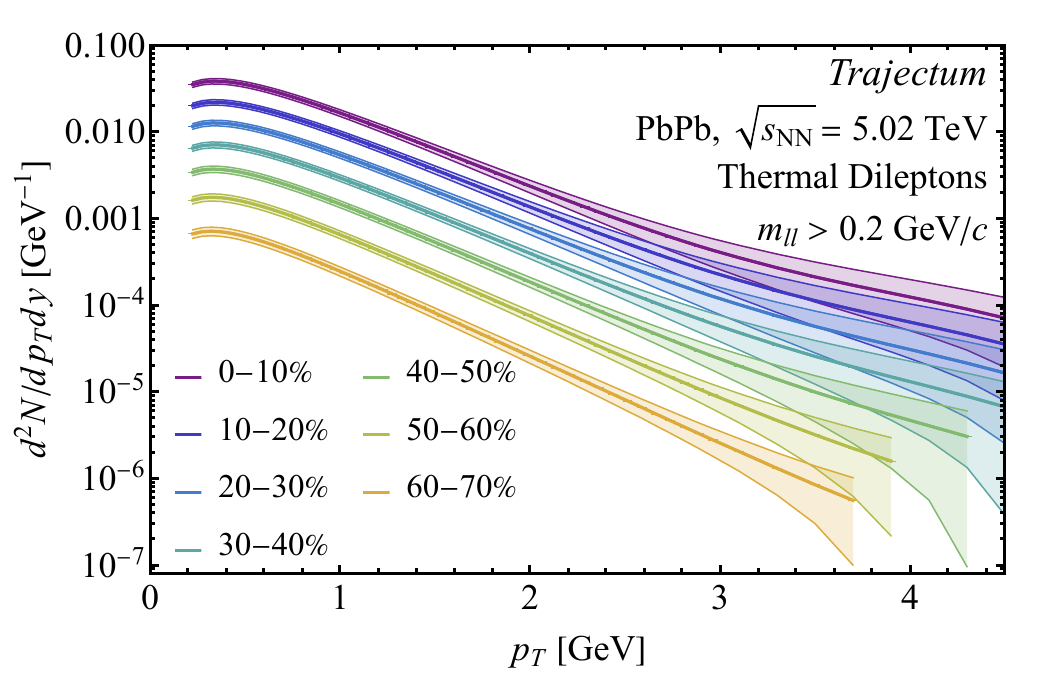}
    \caption{Thermal photon (top) and dilepton (bottom) $p_T$-spectra for 7 different collision centralities from central (0--10\%) to peripheral (60--70\%) collisions. For the thermal photon $p_T$-spectra both the thermal contribution from the QGP and the contribution from the hadron resonance gas (HRG) are shown, with the ratio in the lower panel. The coloured bands correspond to the systematic uncertainty associated with the \emph{Trajectum} model.}
    \label{fig:photondileptonmultiplicity}
\end{figure}

In Fig.~\ref{fig:photondileptonmultiplicity} we present the $p_T$-spectra of the thermal photons and dileptons in seven different centrality classes. The bulk of the hadron resonance gas contribution to the photon yield is concentrated at low momenta ($p_T < 1$\,GeV$/c$), even exceeding the QGP contribution at very low momenta \cite{Huovinen:2001wx}\@. At intermediate and high $p_T$ the contribution of the hadron resonance gas becomes a few percent. As the dilepton rate has an additional factor $\alpha_\text{EM}$, the $p_T$-differential yields of the dileptons are around two to three orders of magnitude smaller compared to the photons. The shape of the dilepton spectra at low-$p_T$ show a maximum around $0.4$\,GeV$/c$, as we consider dileptons with a invariant mass larger than 0.2\,GeV$/c$\@.

\begin{figure}
    \centering
    \includegraphics[width=\columnwidth]{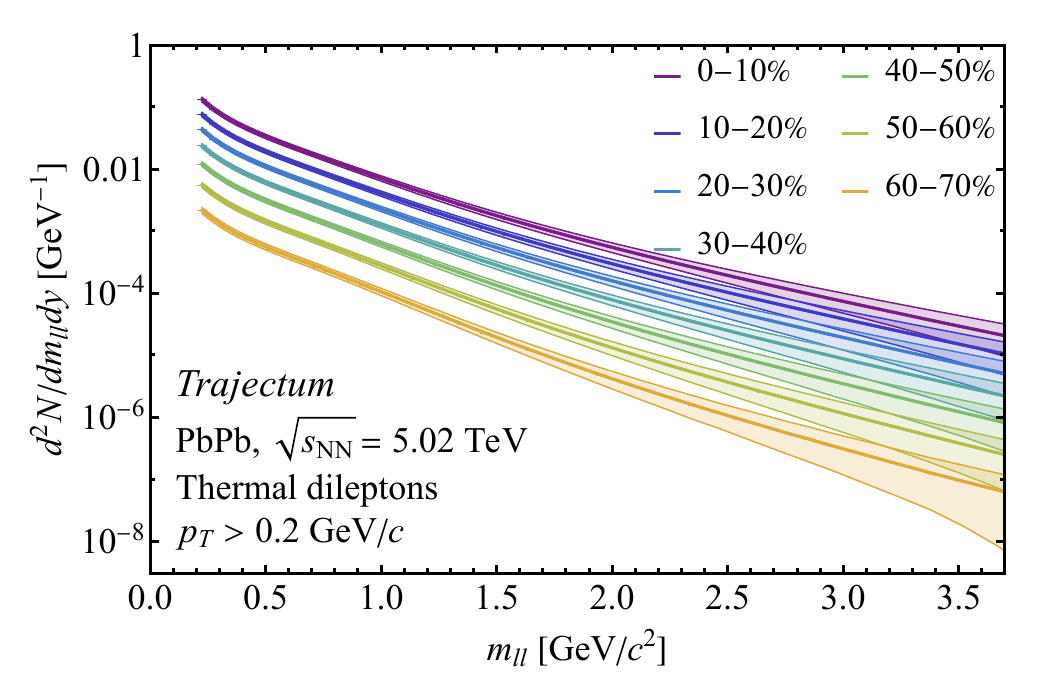}
    \includegraphics[width=\columnwidth]{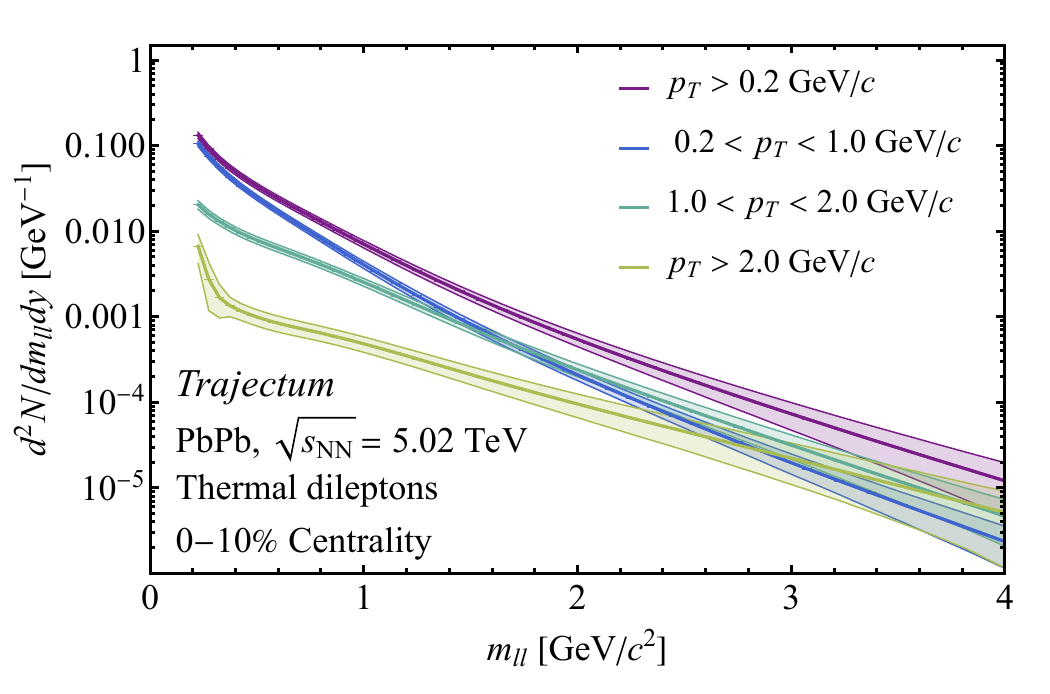}
    \caption{Thermal dilepton invariant mass spectra for 7 different collision centralities (top) from central (0--10\%) to peripheral (60--70\%) Pb--Pb collisions at $\sqrt{s_\text{NN}}=5.02$\,TeV\@ and the result for 0--10\%, differentially for dilepton $p_T$ intervals (bottom)\@. The coloured bands correspond to the systematic uncertainty associated with the \emph{Trajectum} model. 
    }
    \label{fig:dileptonmassspectrum}
\end{figure}

\begin{table*}
\centering
\scalebox{0.99}{
\begin{tabular}{lcccc} \toprule
     & \multicolumn{4}{c}{$\langle \tau \rangle$ [fm/$c$]} \\[0.9ex]
   \text{0--10\% centrality}  & \hspace{4mm}  $p_{T}>0.2$\,GeV$/c$ \hspace{4mm} & $0.2<p_{T}<1.0$\,GeV$/c$ \hspace{4mm} & $1.0<p_{T}<2.0$\,GeV$/c$ \hspace{4mm} & $p_{T}>2.0$\,GeV$/c$ \hspace{4mm} \\[0.9ex]
\toprule
 \textbf{Thermal photons}   & 5.37  & 5.51 &  4.42 &  3.28\\[1ex]
    \midrule
    \textbf{Thermal dileptons} & 4.82 & 3.29 & 3.15 & 2.75 \\[1ex]
    $0.2<m_{ll}<1~$\GeVmass    & 4.97 & 5.19 & 4.43 & 2.99 \\[1ex]
    $1<m_{ll}<2~$\GeVmass      & 3.34 & 3.40 & 3.32 & 3.02 \\[1ex]
    $2<m_{ll}<3~$\GeVmass      & 1.42 & 1.47 & 1.42 & 1.30 \\[1ex]
    $3<m_{ll}<4~$\GeVmass      & 0.91 & 0.96 & 0.94 & 0.83 \\[1ex]
\bottomrule
\end{tabular}
}
\caption{Average emission times ($\langle \tau \rangle$ [fm/$c$]) of thermal photons and dileptons for central (0--10\% centrality)\@ \PbPb{} collisions. The results are given for different selection criteria on transverse momentum $(p_{T})$ of the photons and dileptons, as well as for different intervals in dilepton invariant mass ($m_{ll}$)\@.}
\label{Table:emissiontimes}
\end{table*}

Fig.~\ref{fig:dileptonmassspectrum} shows the invariant mass distributions of dileptons for several collision centralities (top), and the invariant mass spectrum of the dileptons in central collisions for different selections in  transverse momentum (bottom)\@. At small invariant mass ($m_{ll}<1.0$\,GeV$/c^2$), most of the dilepton pairs have $p_T<1.0$\,GeV$/c$\@. In the intermediate mass regime ($1.0 < m_{ll} < 3.0$\,GeV$/c^2$) the contributions of dilepton pairs with $p_T<1.0$\,GeV$/c$ and $1.0<p_T<2.0$\,GeV$/c$ become of the same order of magnitude while at higher invariant masses ($m_{ll}>3.0$\,GeV$/c^2$) the high $p_T$ dileptons start to contribute more significantly. We see that for both the thermal photons and the thermal dileptons the systematic uncertainties start to increase as we move towards higher momenta and higher masses. These high-$p_T$ high mass probes are predominantly emitted from earlier stages of the evolution. As most observables used to constrain the \emph{Trajectum} parameters are from particles produced during late stages of the evolution, the uncertainty on particle production from the early-stages is larger. Furthermore, as our implementation does not include pre-hydrodynamic photon and dilepton production, the dominant parameter contributing to this uncertainty is the starting time of hydrodynamics $\tau_{\rm hyd}$\@. The sensitivity to $\tau_{\rm hyd}$ could be reduced by including off-equilibrium photons and dileptons that are produced before $\tau_{\rm hyd}$ \cite{Garcia-Montero:2023lrd}\@.

It is naively expected that the high-energy thermal emissions happen earlier during the evolution of the QGP, as the system is hotter. This implies that selections on transverse momentum $(p_{T})$ for both photons and dileptons, as well as the invariant mass $(m_{ll})$ in case of dileptons, can be used to discriminate early and late emissions. In Tab.~\ref{Table:emissiontimes} we show the average emission times $(\langle \tau \rangle \, [\mathrm{fm}/c])$ of thermal photons and dileptons, specifically for 0--10\% and 30--40\% collision centralities, which is a sample where the average lifetime of the QGP is about $12$ and $9.5\,\mathrm{fm}/c$ respectively. For both thermal photons and dileptons, it is clear that low-$p_{T}$ emissions originate, on average, from significantly later times in the evolution compared to high-$p_{T}$ emissions. The low momentum emissions happen all throughout the QGP lifetime, and as such the average comes out at about 5--6\,fm$/c$\@. Interestingly, with further selection on the dilepton invariant mass, it is possible to even have $\langle \tau \rangle < 1 \,\mathrm{fm}/c$, beyond what is accessible with thermal photons.

\section{Effective temperatures of the QGP}

\begin{figure}[h!]
    \centering
    \includegraphics[width=\columnwidth]{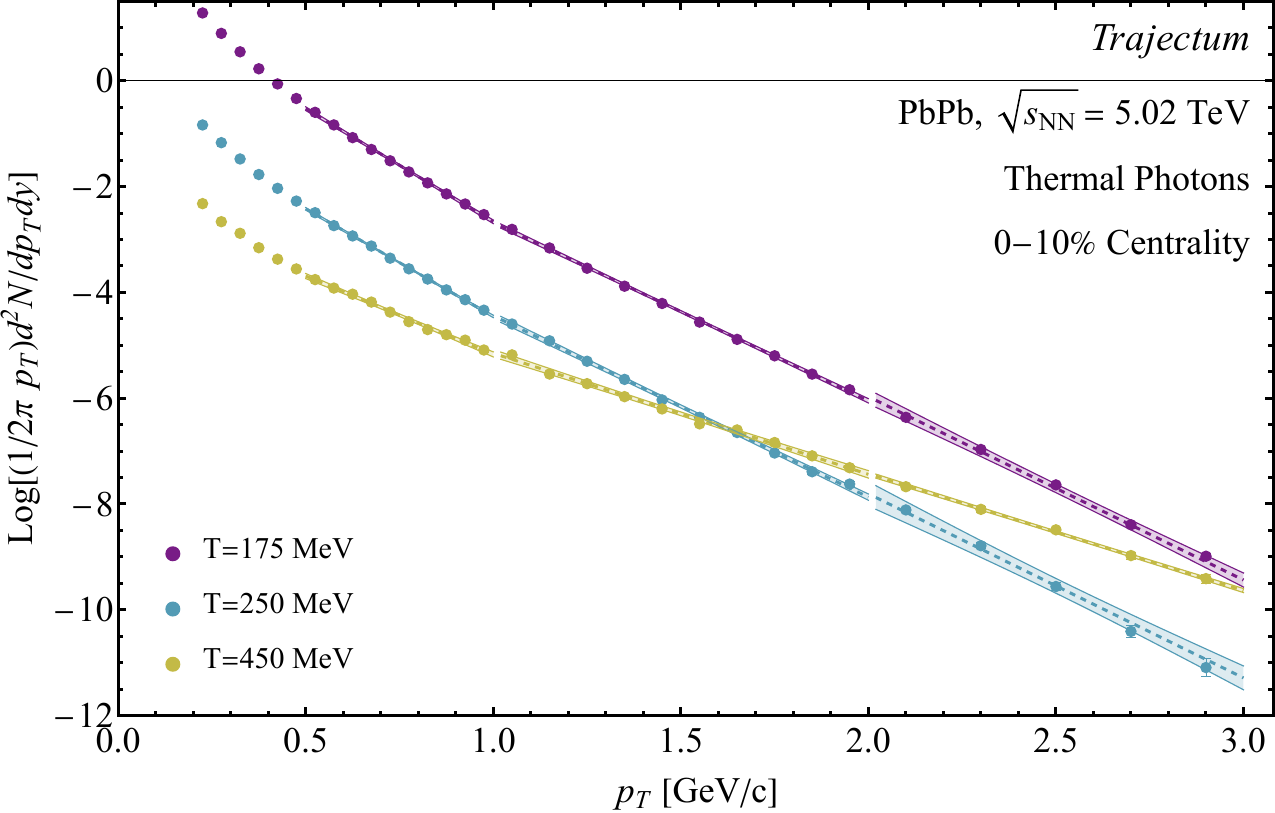}
    \includegraphics[width=\columnwidth]{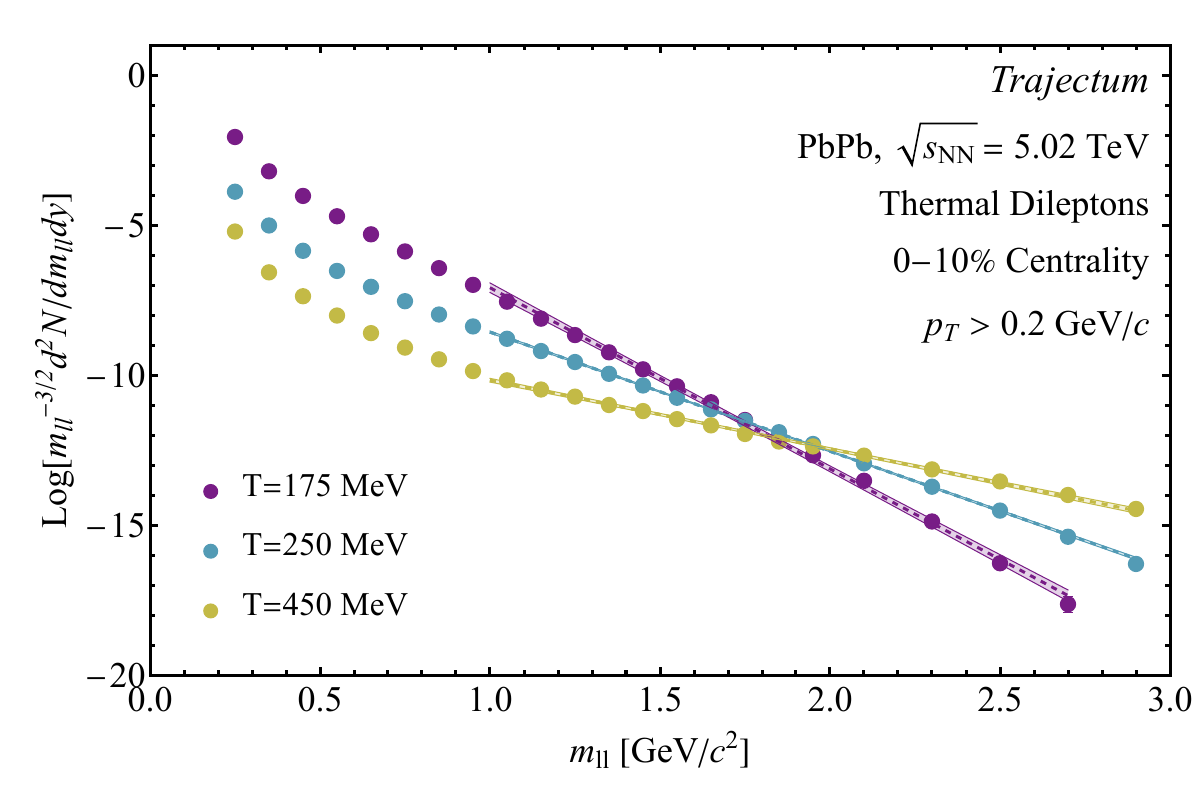}
    \includegraphics[width=\columnwidth]{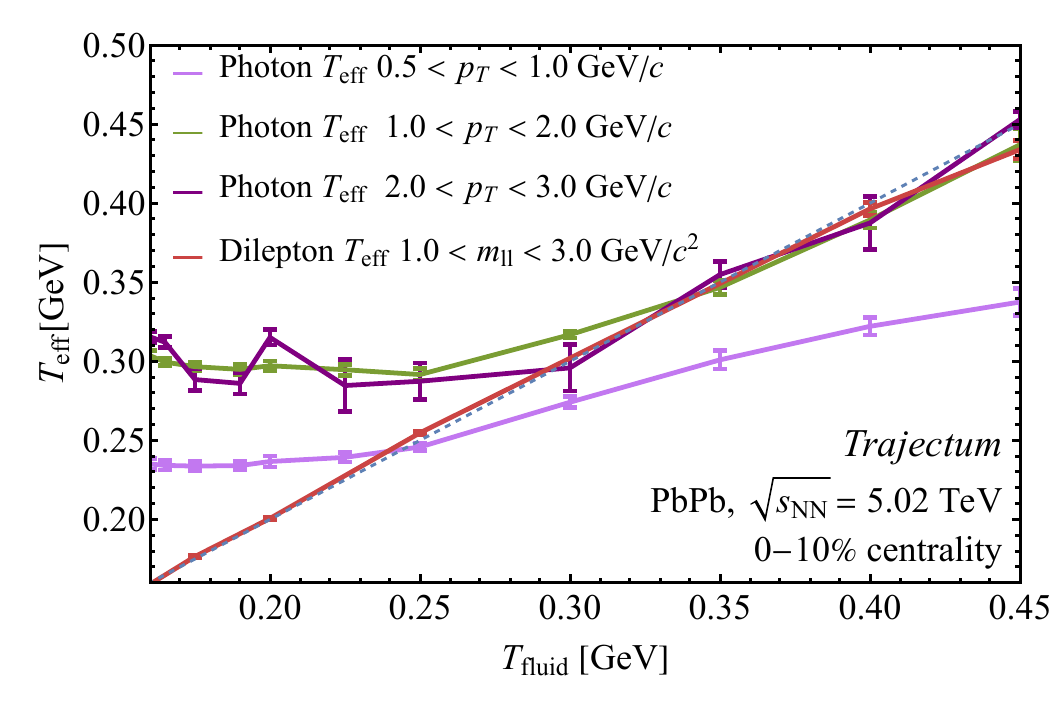}
    \caption{The top and middle panel show the $p_T$ spectra of photons and the invariant mass spectra of dileptons, respectively, emitted from fluid cells at temperatures $T_{\mathrm{fluid}}=175,250\,$and$\,450$\,MeV\@. The effective temperature extraction fit is depicted as a dotted line with a shaded band to indicate the uncertainty associated to the fitting procedure. In the bottom plot we show the effective temperature ($T_{\text{eff}}$) as extracted from the $p_T$-spectra of photons and the invariant mass distribution of dileptons, emitted only from fluid cells at a certain fixed $T_{\mathrm{fluid}}$\@. The diagonal gray dashed line shows $T_{\text{eff}} = T_{\mathrm{fluid}}$\@.}
    \label{fig:effectivetemperaturecheck}
\end{figure}

Measuring the temperature evolution of the QGP is, as explained in the introduction, challenging and can only be indirectly inferred from the properties of the particles that are produced by the fireball. In this work, using the implementation of thermal photons and dileptons into \emph{Trajectum}, we calculate the QGP effective temperatures from their spectra.

In the case of thermal photons, an effective temperature ($T_{\text{eff}}$) can be extracted from the measured $p_T$-spectra. Under the assumption that the production of photons is governed by a Boltzmann distribution one can fit an effective temperature using:
\begin{equation}
\label{eq:TeffPhoton}
    \frac{d^2N}{2\pi p_T\,dp_T\,dy} \propto \exp\left(-\frac{p_T}{T_{\text{eff}}}\right),
\end{equation}
where we require that $p_{T} \gg T$\@. As the $p_T$ of thermal photons is not a Lorentz invariant quantity, the radial flow of the plasma boosts the $p_T$ of the emitted photons towards higher $p_T$, leading to a blue shift in the photon $p_T$-spectra. This shift results in an increase of the effective temperatures.

For thermal dileptons, an effective temperature can be extracted from their invariant mass distribution in the intermediate mass regime (IMR) ($1.0 < m_{ll} < 3.0$ GeV$/c^2$)\@. In this region, where $m_{ll} \gg T$, the emission rate in the local fluid rest frame can be approximated by \cite{Geurts:2022xmk}:
\begin{equation}
\label{eq:TeffDilep}
    \frac{d^2N}{dm_{ll}\,dy} \propto (m_{ll}T_{\text{eff}})^{3/2}\exp\left(-\frac{m_{ll}}{T_{\text{eff}}}\right).
\end{equation}

The invariant mass of the dilepton pair is a Lorentz invariant quantity and consequently does not suffer from a blue shift, and it was indeed shown by model calculations that dileptons are a good probe for the temperature of the QGP \cite{Churchill:2023vpt, Churchill:2023zkk}\@. In the following paragraph we show that this holds true in our framework as well.

In Fig.~\ref{fig:effectivetemperaturecheck} we study the correspondence between the effective temperature ($T_{\text{eff}}$) and the temperature of the emitting fluid cell ($T_{\mathrm{fluid}}$)\@. Fig.~\ref{fig:effectivetemperaturecheck} top and middle show the fits of the effective temperature of the thermal photons and dileptons for $T = 175,\,250$ and $450\,$MeV, respectively. These fits were done on the equal weighted logarithm of the spectrum in the ranges $0.5 < p_{T} < 1.0$\,GeV$/c$, $1.0 < p_{T} < 2.0$\,GeV$/c$, and $2.0 < p_{T} < 3.0$\,GeV$/c$ for thermal photons, and in the IMR ($1.0 < m_{ll} < 3.0$\,GeV$/c^2$) for thermal dileptons.

It is interesting that below an invariant mass of around $m_{ll} = 1.8$\,GeV$/c^2$ the spectrum is dominated by lower temperatures. This can be explained by the competition between the lifetime of the plasma, which spends more time at lower temperatures, and the fact that high invariant mass dileptons require a higher temperature. In Fig.~\ref{fig:effectivetemperaturecheck} (bottom) we show the effective temperature of the fit versus the fluid temperature where the fit was performed for both the dileptons as well as fits to the photon spectrum in the three $p_T$ ranges mentioned above. These ranges are similar to the range used in most experiments. The effective temperature as extracted from the thermal dilepton invariant mass distribution follows the emission temperature closely over the entire range of $T_{\rm fluid}$ as also seen in \cite{Churchill:2023vpt}\@. The thermal photon $T_{\text{eff}}$ for $1.0< p_{T} < 2.0$\,GeV$/c$\@ and $2.0< p_{T} < 3.0$\,GeV$/c$\@, on the other hand, shows a clear deviation at lower $T_{\mathrm{fluid}}$, indicating that dileptons are better suited to relate their effective temperature to the temperature of the plasma\@. This is explained by the interplay of the temperature, emission times, and the build-up of radial flow. The higher the radial flow, the more the thermal photon blue-shifts in $p_T$\@. This leads to an effective temperature for the photons of around $T_{\rm eff} \sim 250$--300\,MeV\@. For the lower $p_T$ photons, from $0.5 < p_{T} < 1.0$\,GeV$/c$\@, we see a blueshift for fluid temperatures below $250$\,MeV, but for higher values it shows $T_{\rm eff} < T_{\rm fluid}$. For these fluid temperatures we are no longer in the regime that $p_{T} \gg T$, as required for the Boltzmann approximation. This is also reflected in the top of Fig.~\ref{fig:effectivetemperaturecheck}, where you can see a clear deviation from an exponential for the low-momentum photons.

\begin{figure}[h]
    \centering
    \includegraphics[width=\columnwidth]{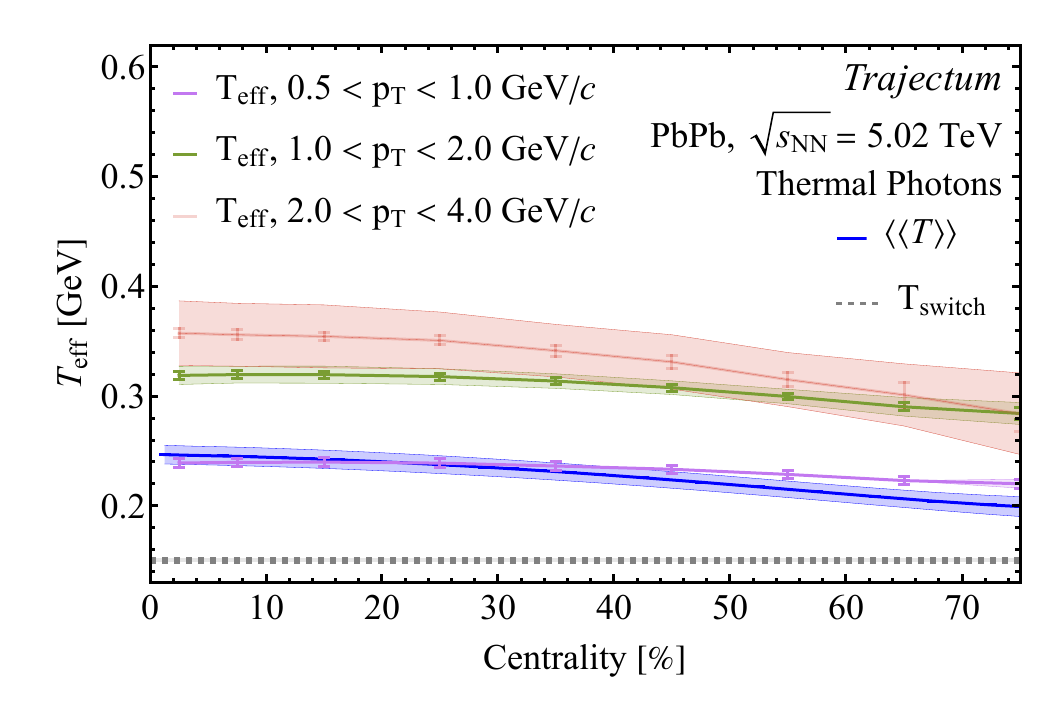}
    \includegraphics[width=\columnwidth]{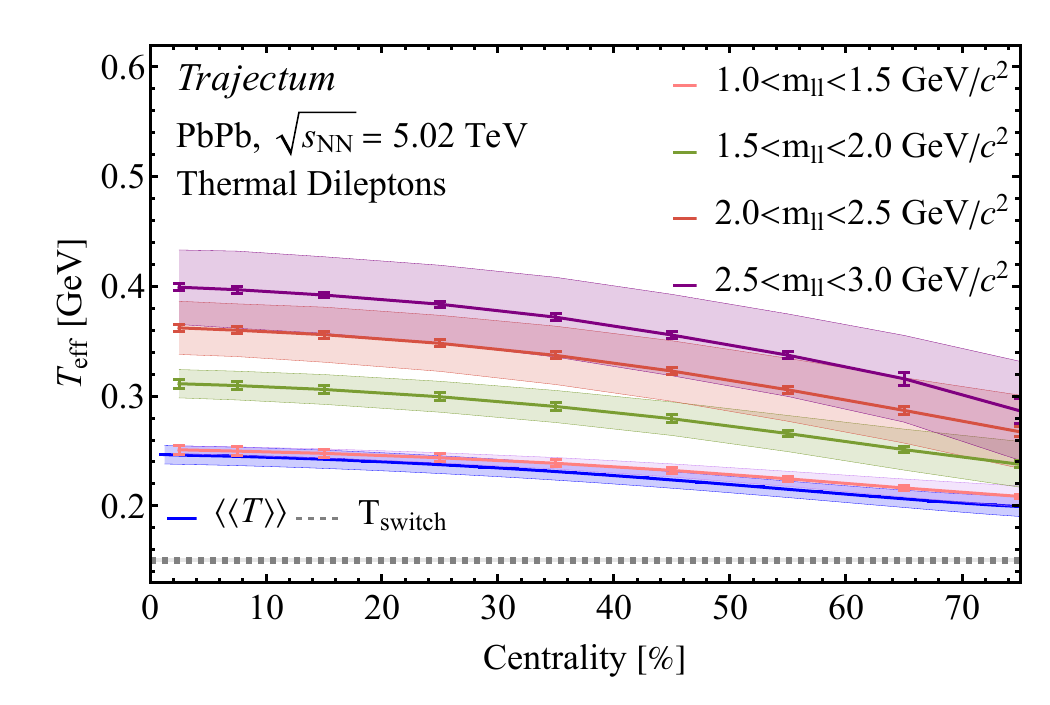}
    \caption{The thermal photon (top) and dilepton (bottom) effective temperatures as a function of collision centrality for different fit ranges in $p_{T}$ and $m_{ll}$, respectively. The effective temperatures are compared to the time-averaged average temperature of the fluid cells ($\langle \langle T \rangle \rangle$)\@. The coloured bands correspond to the systematic uncertainty associated with the \emph{Trajectum} model while the markers correspond to the uncertainty associated with the fitting procedure.}
    \label{fig:effectivetemperature}
\end{figure}

Fig.~\ref{fig:effectivetemperature} presents the effective temperatures of thermal photons (top) and dileptons (bottom), as a function of the collision centrality, together with the space- and time-averaged temperature. The functional forms of Eqs.~(\ref{eq:TeffPhoton}--\ref{eq:TeffDilep}) are used to fit the respective $p_{T}$ and $m_{ll}$ distributions. These fits are performed in three different $p_{T}$ regions and 4 different $m_{ll}$ regions for the photons and dileptons respectively. The bands show the systematic uncertainty on the effective temperature whereas the fences on the plot markers show the combined fitting uncertainty and statistical uncertainty. For both the thermal photons and dileptons, a slight centrality dependence of the effective temperature is observed, albeit more subtle for the photons.

In central collisions the effective temperature of both probes is around $T_{\text{eff}} \sim 250$--300\,MeV\@. For the thermal photons this effective temperature is an interplay between the temperature of the plasma and the blue-shift experienced by the photons. In Fig.~\ref{fig:effectivetemperaturecheck} it was shown that photons emitted from the plasma at $T<250$\,MeV have $T_{\text{eff}} \sim 250$--300\,MeV\@. This temperature range is also responsible for the bulk of the thermal photon production, whereas the higher temperature fluid cells provide a negligible contribution to the total yield. Towards more peripheral collisions the effective temperature of the plasma decreases, mainly due to the decrease in radial flow.

In case of the thermal dileptons, the effective temperature is similar to the $T_{\text{eff}}$ extracted from the photon spectra, with a similar centrality dependence. For the thermal dilepton $T_{\text{eff}}$, the higher effective temperature is caused by the bias towards earlier emissions and consequently higher plasma temperatures, as can be seen in Tab.~\ref{Table:emissiontimes}\@. In contrast to the photons, the bulk of the emission of dileptons in the IMR does not occur during the later stages when the system is at a lower temperature. In the IMR all temperatures provide a sizeable contribution to the total production, where the lower temperatures dominate below $m_{ll} = 1.8$\,GeV$/c^2$ and vice versa, as can be seen in the top of Fig.~\ref{fig:effectivetemperaturecheck}\@. The effective temperatures in the different invariant mass bins show exactly the expected behaviour. The extracted effective temperature on the lower end of the IMR ($1 < m_{ll} < 1.5$\,GeV$/c^2$) is lower than the extracted temperature on the higher end of the IMR ($2.5 < m_{ll} < 3$\,GeV$/c^2$)\@. Note that pre-equilibrium as well as Drell-Yan contributions start to play an important role at higher invariant masses, which are not included in this study \cite{Coquet:2021lca}\@. In addition, at lower invariant masses the hadronic contributions dominate over the QGP contributions \cite{Bratkovskaya:2014mva}\@.

\begin{figure}
    \centering
    \includegraphics[width=\columnwidth]{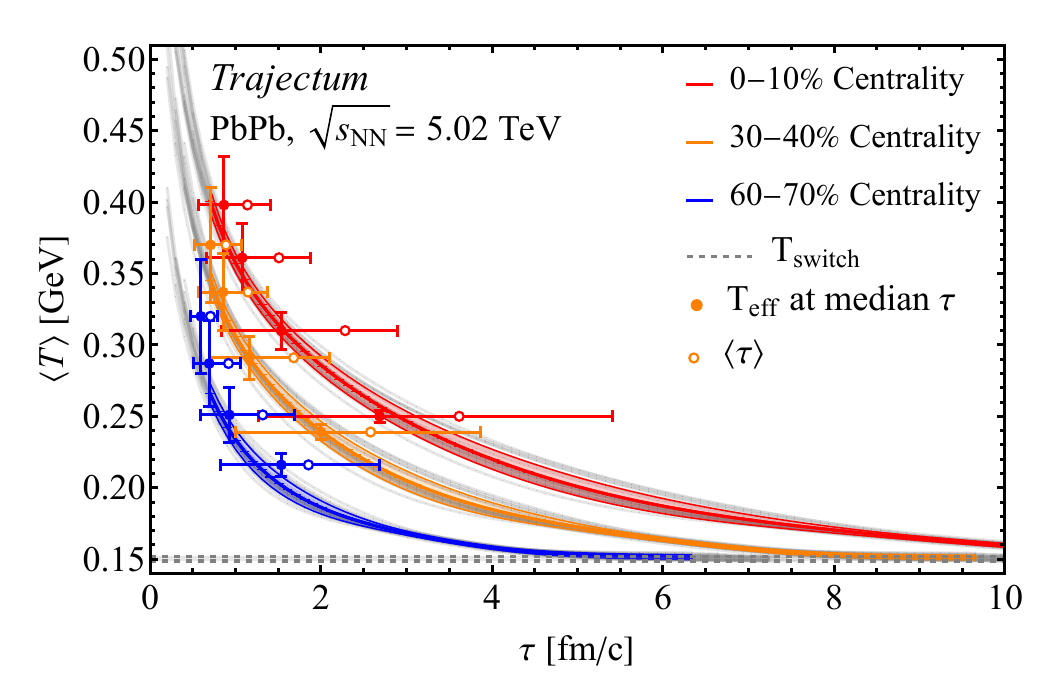}
    \caption{
    Comparison of $\langle T \rangle$ as function of $\tau$ to the dilepton $T_{\rm eff}$ and $\tau$ for four equal $m_{ll}$ bins in the IMR given selection criteria. The horizontal width of the markers denotes the first and third quartile of the $\tau$ distribution, with the filled and open marker indicating the median and mean, respectively. The vertical uncertainty corresponds to the systematic uncertainty of $T_{\rm eff}$\@. The 20 parameter variations are shown in gray.
    }
    \label{fig:CompareDileptonQGPtemperature}
\end{figure}

A particularly interesting idea is presented in Fig.~\ref{fig:CompareDileptonQGPtemperature}\@. For three centralities we take dileptons with $m_{ll}$ binned in four equal bins in the IMR from $1$ to $3$\,GeV$/c^2$ and show their effective temperature and production time. The production time comes from \emph{Trajectum} and is a relatively wide distribution represented by the median (solid), mean (open) and the first and third quartiles. Nevertheless, as also seen in Tab.~\ref{Table:emissiontimes}, higher $m_{ll}$ dileptons are emitted significantly earlier. It is remarkable that the effective temperatures (which are now experimentally accessible) agree to an excellent degree with the \emph{Trajectum} temperature presented earlier in Fig.~\ref{fig:temperaturesTau}, provided we take the temperature at the median time of the emitted dileptons. This is true across a wide range of centralities, dilepton masses and effective temperatures.

\section{Anisotropic flow}

\begin{figure}
    \centering
    \includegraphics[width=\columnwidth]{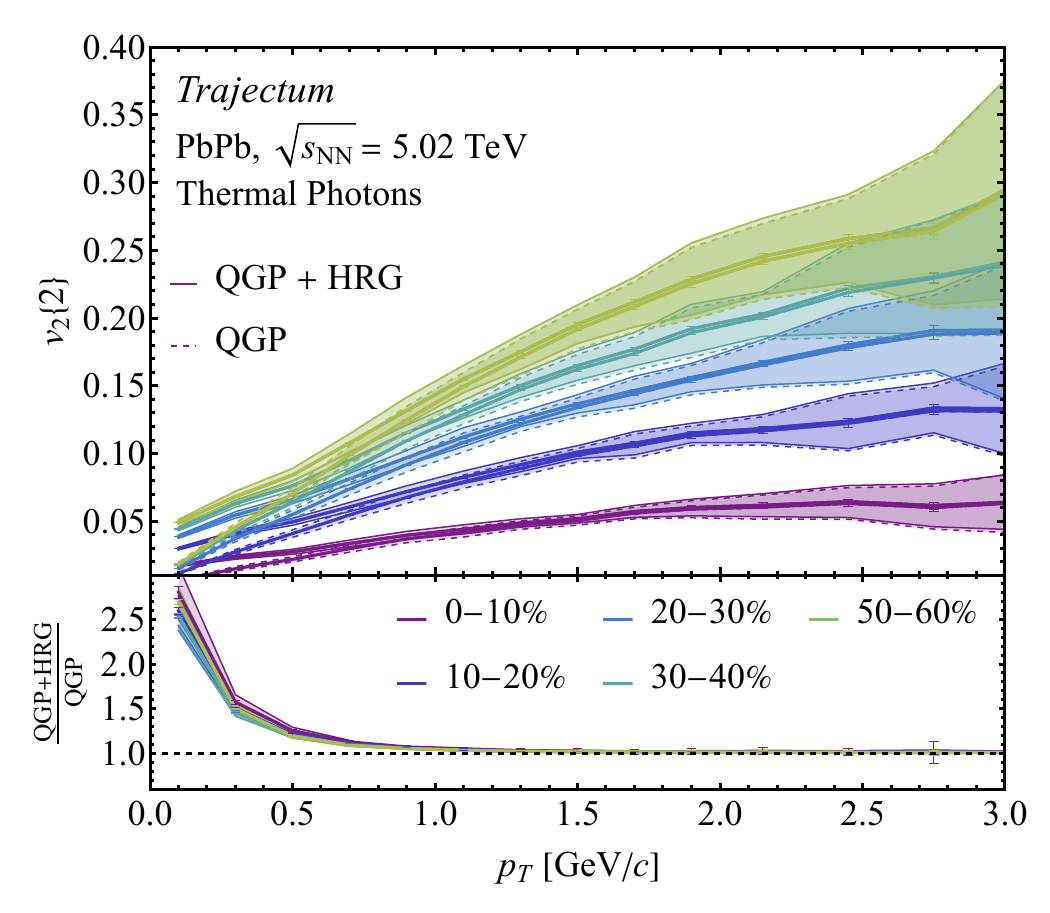}
    \includegraphics[width=\columnwidth]{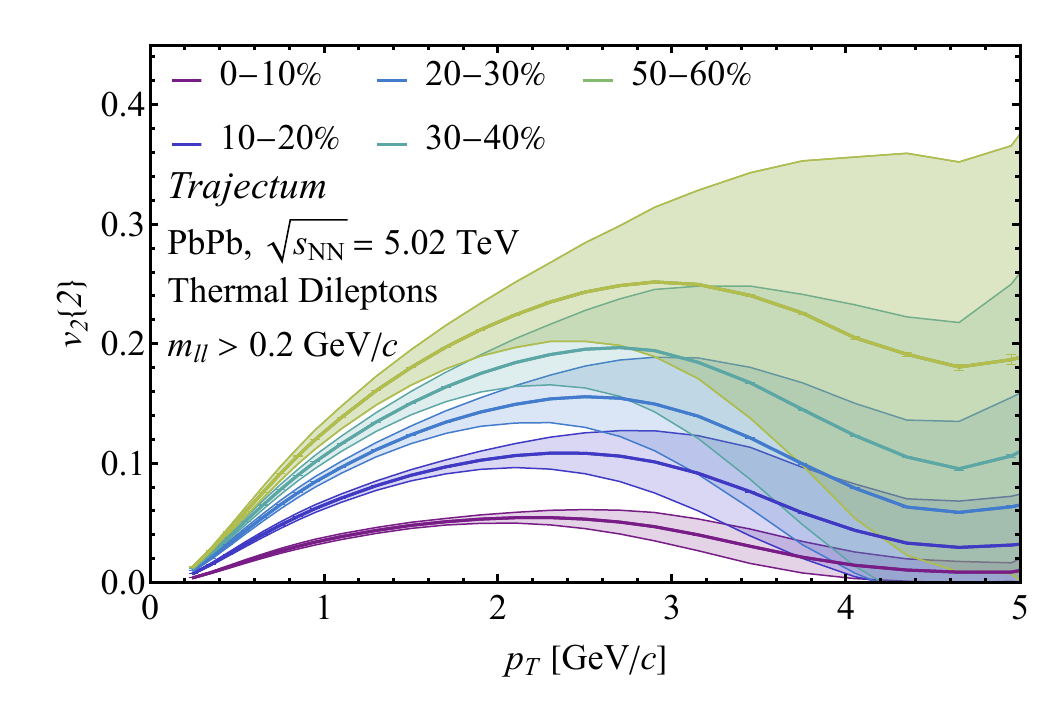}
    \caption{The elliptic flow of thermal photons (top) and thermal dileptons (bottom) as a function of the transverse momentum. We show the elliptic flow for 5 different collision centralities from central (0--10\%) to semi-peripheral (50--60\%) collisions. For the thermal photons we show both the QGP contribution and the contribution from the hadron resonance gas (HRG) and the QGP combined, with the ratio in the lower panel. The coloured bands correspond to the systematic uncertainty associated with the \emph{Trajectum} model.}
    \label{fig:v2vspT}
\end{figure}

In addition to the spectra of the thermal photons and dileptons, their anisotropic flow provides an important additional probe of the time dependent temperature and collective behaviour of the QGP\@.
In this work we have determined the anisotropic flow of the thermal photons and dileptons using the cumulant method \cite{Bilandzic:2010jr, Bilandzic:2012wva}\@. We use the same method as is prevalent in experimental studies, to provide an apples to apples comparison. As discussed in the previous sections it is possible to single out certain times in the evolution of the system by applying certain selection criteria on the photons or dileptons. In this section we will see if, by considering the elliptic flow in a certain mass or transverse momentum range, we can obtain the elliptic flow of the system at earlier or later times. We do this by determining the elliptic flow with different selections on the production time $\tau$ of the photon or dilepton pair, which is, in contrast to experiments, available in \emph{Trajectum}\@.

\begin{figure}
    \centering
    \includegraphics[width=\columnwidth]{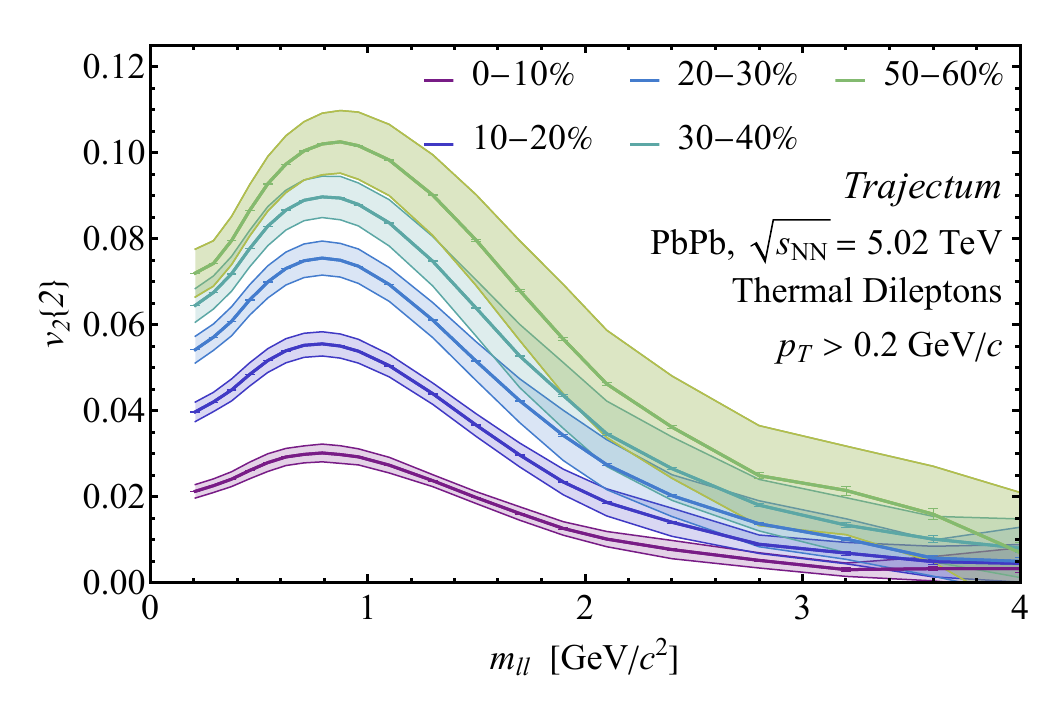}
    \includegraphics[width=\columnwidth]{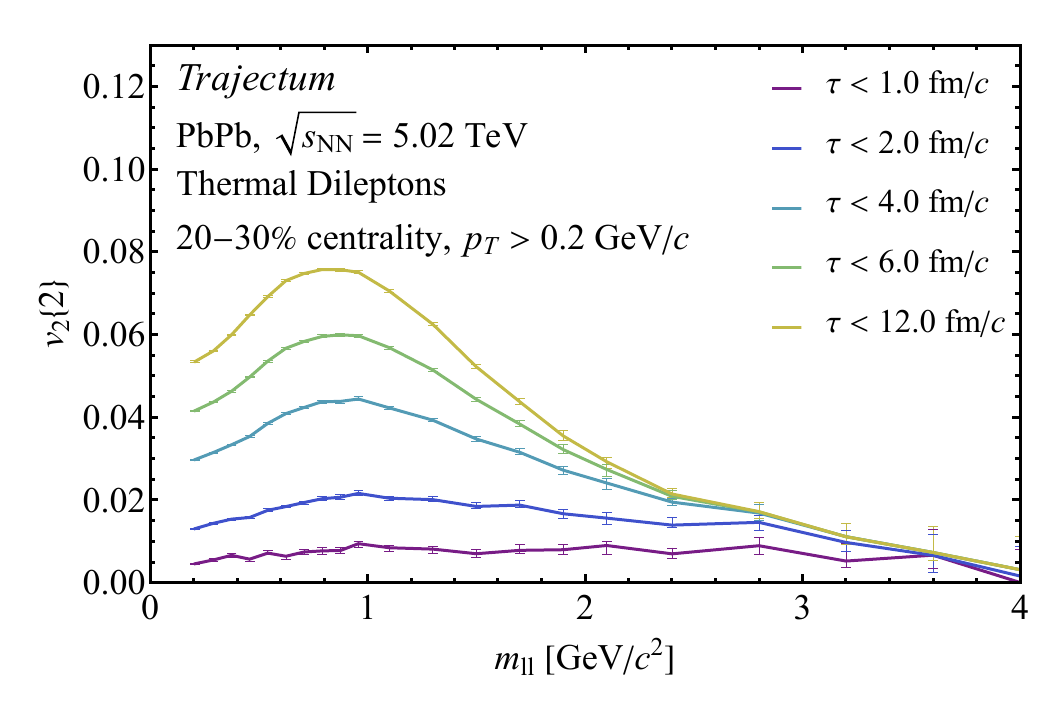}
    \caption{The elliptic flow of dileptons as a function of the invariant mass of the dilepton pair. In the top panel we show the elliptic flow for 5 different collision centralities from central (0--10\%) to semi-peripheral (50--60\%) collisions. The coloured bands correspond to the systematic uncertainty associated with the \emph{Trajectum} model. In the bottom panel we show how the elliptic flow develops if we consider dileptons produced at production times $\tau$\@.} 
    \label{fig:dileptonv2vsQ2}
\end{figure}

In the top panel of Fig.~\ref{fig:v2vspT} we present the elliptic flow of thermal photons for five different centrality classes, again separating the total thermal and the pure QGP contributions. As in the $p_T$-spectrum the HRG photons have a significant contribution to the total elliptic flow at low transverse momentum, while at higher momenta their contribution becomes negligible. In the bottom of Fig.~\ref{fig:v2vspT} and in Fig.~\ref{fig:dileptonv2vsQ2} we present the elliptic flow of thermal dileptons as a function of transverse momentum and invariant mass, respectively, in the same five centrality classes. As expected, we see an increase in the elliptic flow for more peripheral collisions combined with an increase for intermediate momenta. For the highest momenta we see a decrease in the flow. These highest momentum dileptons are predominantly produced at the early stages of the collision when the flow has not yet developed. In Fig.~\ref{fig:dileptonv2vsQ2} we observe an increase in the elliptic flow as the invariant mass increases, peaking at an invariant mass of approximately $800$\,MeV$/c^2$\@. As the invariant mass of the dileptons increases, their mean transverse momentum also increases, in turn increasing the flow. However, we have seen that the invariant mass of the dilepton pair is an even stronger indicator of the production time. The high mass dilepton pairs stem from the earliest times of our system and therefore the flow should go towards 0 for these high masses, which is what we observe.

In the bottom panel of Fig.~\ref{fig:dileptonv2vsQ2}, we study the elliptic flow as a function of the invariant mass of the dilepton pair for different constraints on the production time $\tau$ of the dilepton pair. While the flow in the low-mass regime develops gradually over time, for higher masses ($m_{ll} > 2$\,GeV$/c^2$) the main contribution to the flow stems from early time dileptons ($\tau < 4$\,fm$/c$)\@. This behaviour is again consistent with the results from Tab.~\ref{Table:emissiontimes}, where we have seen that the invariant mass of the dilepton pair is a good discriminator for the production time, even more so than the transverse momentum. In addition it confirms our explanation of the shape of the flow as a function of the invariant mass given above.

Another observation is an increase in the systematic uncertainties in the elliptic flow of both the photons and the dileptons as we go towards higher masses and momenta, as we have seen for the spectra as well. In addition the systematic uncertainty also increases for more peripheral collisions. In the case of the dileptons we know that the high-mass high-momentum pairs are predominantly created during the early stages of the evolution of the QGP\@. However, the observables used to constrain the \emph{Trajectum} parameters in the Bayesian analyses are dominated by the later stages of the QGP evolution.

\section{Conclusions and outlook}

We have implemented next-to-leading order thermal emission of photons and dileptons by a weakly coupled plasma into \emph{Trajectum}, a state-of-the-art hybrid model for heavy-ion collisions. In addition we determined systematic uncertainties for all observables, which are calculated using the posterior distributions of the \emph{Trajectum} model parameters. The resulting $p_T$-spectrum of the thermal photons and both the $p_T$-spectrum and invariant mass distribution of the thermal dileptons have been studied. From these spectra the effective temperatures were extracted. The effective temperature of the thermal photons is dominated by the radial flow in the system, and therefore shows little sensitivity to the true temperature of the QGP, where the effective temperatures will be higher than the plasma itself $(T_{\mathrm{eff}}>T_{\mathrm{fluid}})$\@. This implies that measuring a QGP $T_{\mathrm{eff}}$ above the crossover temperature with thermal photons does not guarantee that the actual temperature of the fluid reached that same value. Our model calculations of the effective temperature of thermal photons agree with the experimental measurements performed in recent years, where $T_{\rm eff} \sim 250$--300\,MeV independent of collision centrality and even centre-of-mass energy \cite{ALICE:2015xmh, STAR:2016use, PHENIX:2022rsx, PHENIX:2022qfp, ALICE:2023jef}\@. On the other hand, the effective temperature as extracted from the invariant mass distribution of the dileptons is not affected by radial flow and is an integral over the temperature evolution of the collision system. This means that thermal dileptons can be used to show that the temperature of the fluid exceeds the $T_\text{co}$ predicted by lattice QCD\@.

In addition, we have showcased in Fig.~\ref{fig:effectivetemperature} that by applying different selection criteria on the transverse momentum and the invariant mass it is possible to discriminate between different average emission times ranging from late emission ($\langle \tau \rangle = 5.6\,$fm$/c$) to very early emissions ($\langle \tau \rangle < 1.0\,$fm$/c$)\@. Extracting the effective temperatures from the invariant mass spectrum in the intermediate mass regime allows us to access information on the average temperature at different stages of the collision, where the effective temperature corresponds closely to the average plasma temperature.

Finally we have presented the elliptic flow of thermal photons and dileptons, calculated using the flow-vector cumulants method. Utilising selection criteria on the dilepton pair, especially on the invariant mass, allows us to determine the amount of anisotropic flow at different stages of the system's evolution. The systematic uncertainty is largest in the high-mass high-$p_T$ domain, a domain sensitive to the early stages of the collisions. A precise measurement of the thermal dilepton spectra and elliptic flow will, in particular, constrain the parameters governing the initial stages.

In the appendix we provide a comparison of the spectra and elliptic flow to the ALICE measurement \cite{ALICE:2015xmh} and the previous theoretical studies \cite{vanHees:2014ida, Linnyk:2015tha, Paquet:2015lta,Kasmaei:2018oag, Garcia-Montero:2024lbl}\@. The addition of thermal probes to \emph{Trajectum} marks a starting point for further studies on electromagnetic radiation in heavy ion collisions. Several elements can still be added, including prompt production \cite{vanHees:2014ida, Linnyk:2015tha, Paquet:2015lta}, non-equilibrium production \cite{Garcia-Montero:2024lbl} and viscous effects \cite{Kasmaei:2018oag}\@. Future observables to study include e.g.~dilepton polarisation~\cite{Seck:2023oyt} on an event-by-event basis which can constrain the amount of thermalisation of the QGP \cite{Coquet:2021gms, Coquet:2023wjk}\@.

\clearpage

\bibliographystyle{apsrev4-1}
\bibliography{main, prcmanual}

\section*{Appendix: Comparison to other theoretical calculations}

In this appendix we show how the results presented in this work compare with other theoretical and experimental work. As most previous studies were performed at a collision energy of \Btwonn, we generated events with \textit{Trajectum} for both \Btwonn{} and \Bfivenn{} using maximum a posteriori (MAP) settings, and rescaled all observables shown in this work to provide a one-to-one comparison.

\begin{figure}
    \centering
    \includegraphics[width=\columnwidth]{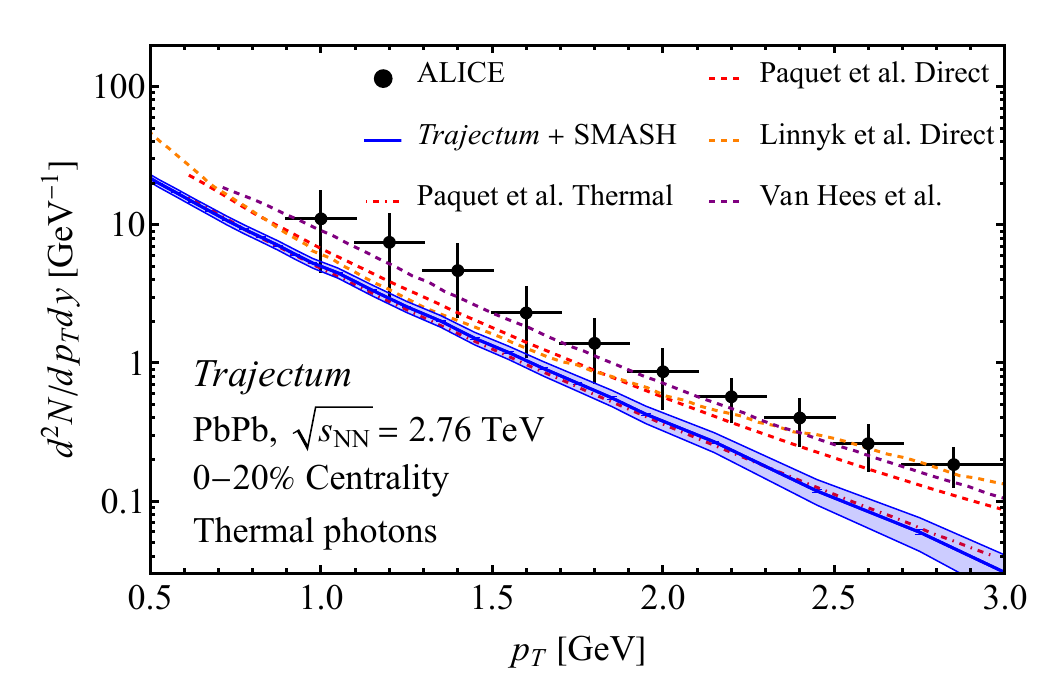}
    \includegraphics[width=\columnwidth]{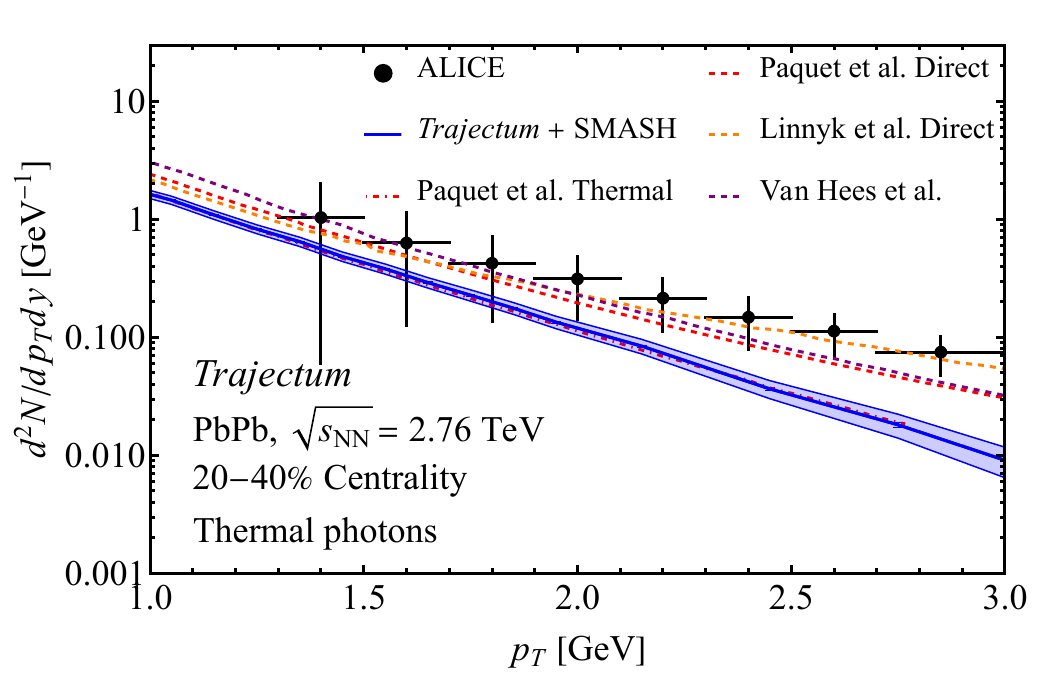}
    \caption{The \textit{Trajectum}  results on the $p_T$-spectrum of thermal photons compared to different theoretical calculations from Paquet et al \cite{Paquet:2015lta} (thermal and direct), Linnyk et al \cite{Linnyk:2015tha} (direct), van Hees et al \cite{vanHees:2014ida} (direct) and the direct photon spectrum measured by ALICE \cite{ALICE:2015xmh}\@, in the 0--20\% (top) and the 20--40\% (bottom) collision centrality interval.}
    \label{fig:theorycompphotonspectra}
\end{figure}

The inclusive photon spectrum can be divided into decay photons coming mostly from $\pi^0$ decays and direct photons coming directly from the collision. Within the direct photons a further separation can be made between photons that are produced thermally or prompt photons from QCD processes. Fig.~\ref{fig:theorycompphotonspectra} compares the thermal photon $p_T$-spectra from this work to the calculations from \cite{vanHees:2014ida, Linnyk:2015tha, Paquet:2015lta}, as well as the experimental data from ALICE \cite{ALICE:2015xmh}\@. The top and bottom panel show the results for the 0--20\% and the 20--40\% collision centrality intervals, respectively. We will now discuss the various similarities and differences between these theoretical calculations.

\begin{figure}
    \centering
    \includegraphics[width=\columnwidth]{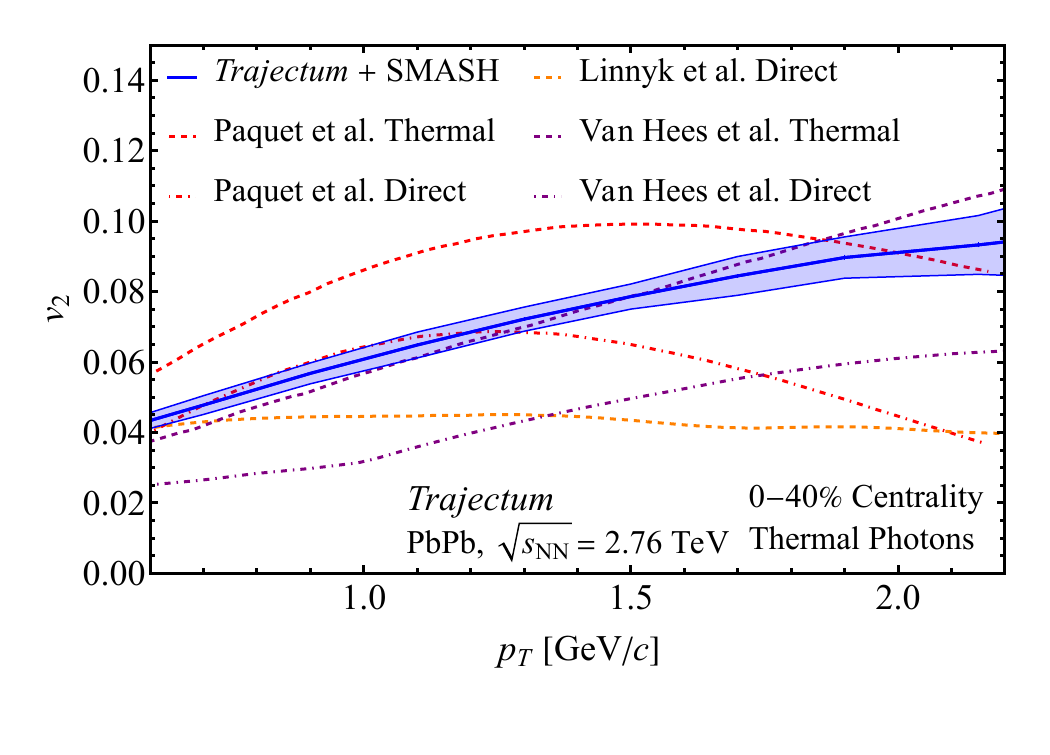}
    \caption{The \textit{Trajectum} results on the elliptic flow of thermal photons compared to different theoretical calculations from \cite{vanHees:2014ida, Linnyk:2015tha, Paquet:2015lta}, in the 0--40\% collision centrality interval.}
    \label{fig:theorycompphotonflow}
\end{figure}

The approach in \cite{Paquet:2015lta} closely resembles the approach in this work. It provides a calculation of both thermal and direct photons in a hybrid model for heavy-ion collisions, using an IP-Glasma initial state, viscous hydrodynamic fluid calculations in 2+1D for the QGP using MUSIC, and UrQMD as a hadronic afterburner. Reference \cite{Paquet:2015lta}  uses the LO rates from \cite{Arnold:2001ba}, while here we use the NLO rates from \cite{Ghiglieri:2014kma}\@. On the other hand, \cite{Paquet:2015lta} includes viscous corrections and hadronic decays beyond the standard cocktail. In addition to the thermal production they have studied the prompt production of photons using pQCD calculations tuned to proton-proton collisions scaled with the number of binary collisions on an event-by-event basis. The comparison to their result is made with both their total photon production, labelled `Direct', and their purely thermal production. We can see in Fig.~\ref{fig:theorycompphotonspectra} that within the systematic uncertainty our yield is in good agreement compared to the thermal component of \cite{Paquet:2015lta}\@. The inclusion of prompt photons results in an increased yield, especially at larger $p_T$, which is expected as the prompt spectrum only decreases as a power law.

\begin{figure}
    \centering
    \includegraphics[width=\columnwidth]{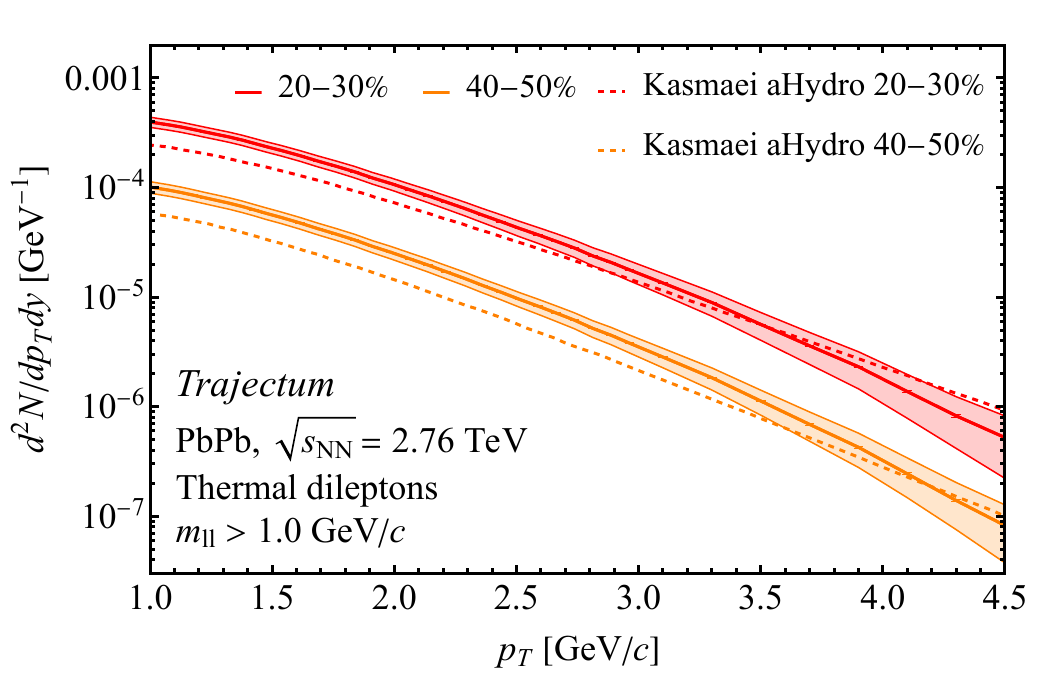}
    \includegraphics[width=\columnwidth]{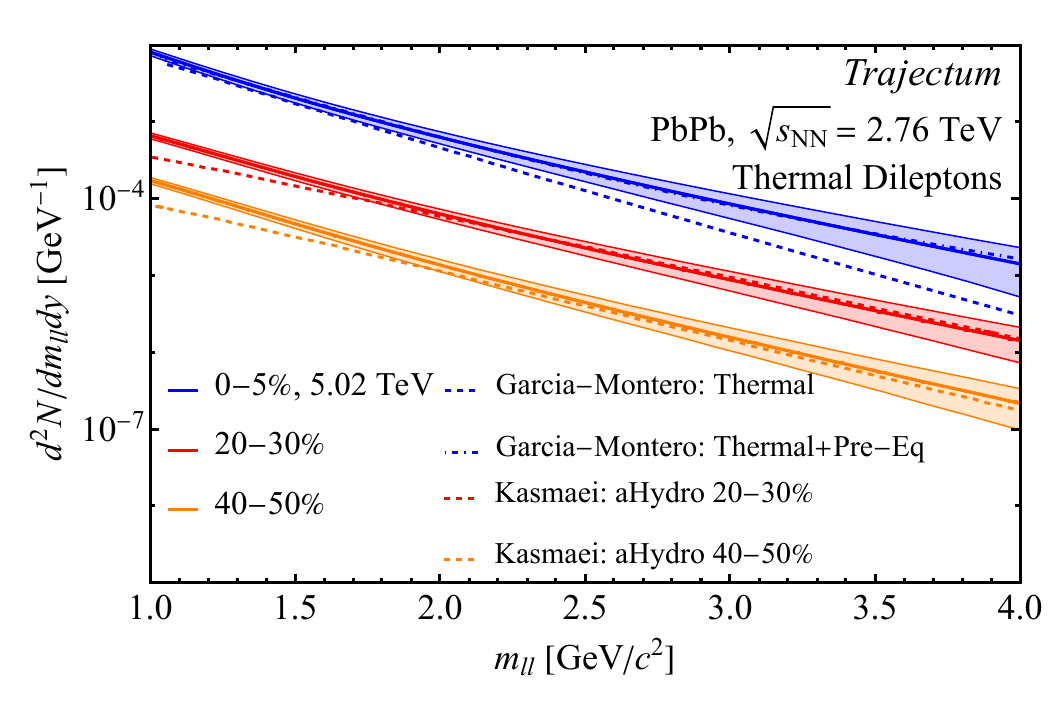}
    \caption{The \textit{Trajectum} results on the $p_T$-spectrum (top) and invariant mass spectrum (bottom) of thermal dileptons compared to \cite{Kasmaei:2018oag, Garcia-Montero:2024lbl}\@, where the kinematic cuts for the results of this work are adjusted such that they match.}
    \label{fig:theorycompdileptonspectra}
\end{figure}

\begin{figure}
    \centering
    \includegraphics[width=\columnwidth]{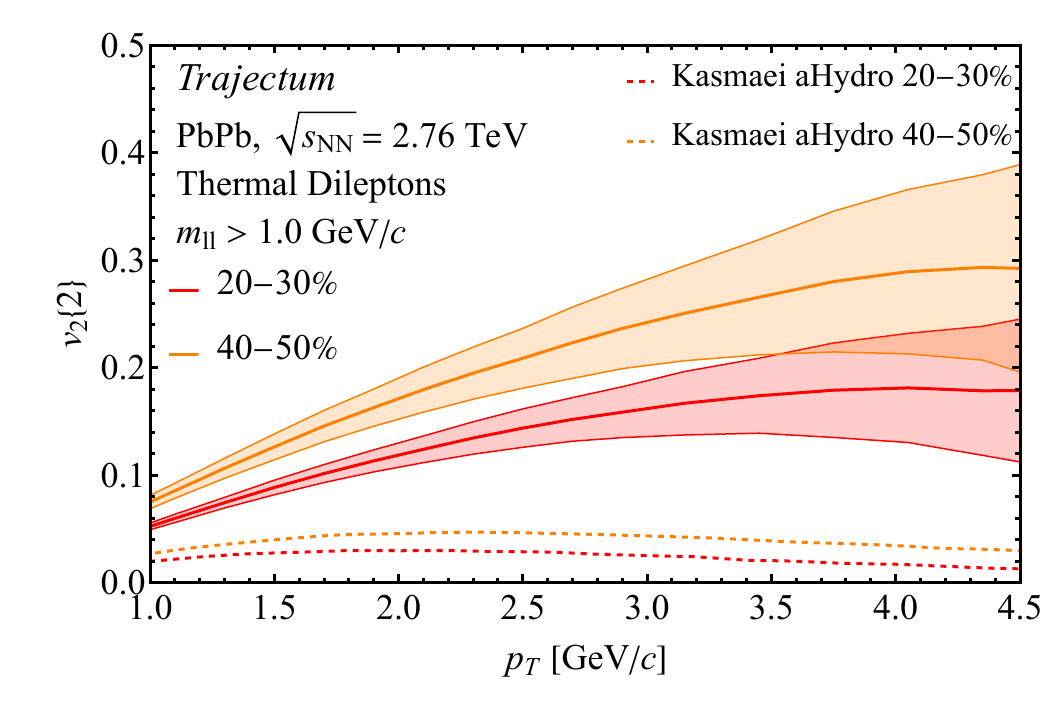}
    \includegraphics[width=\columnwidth]{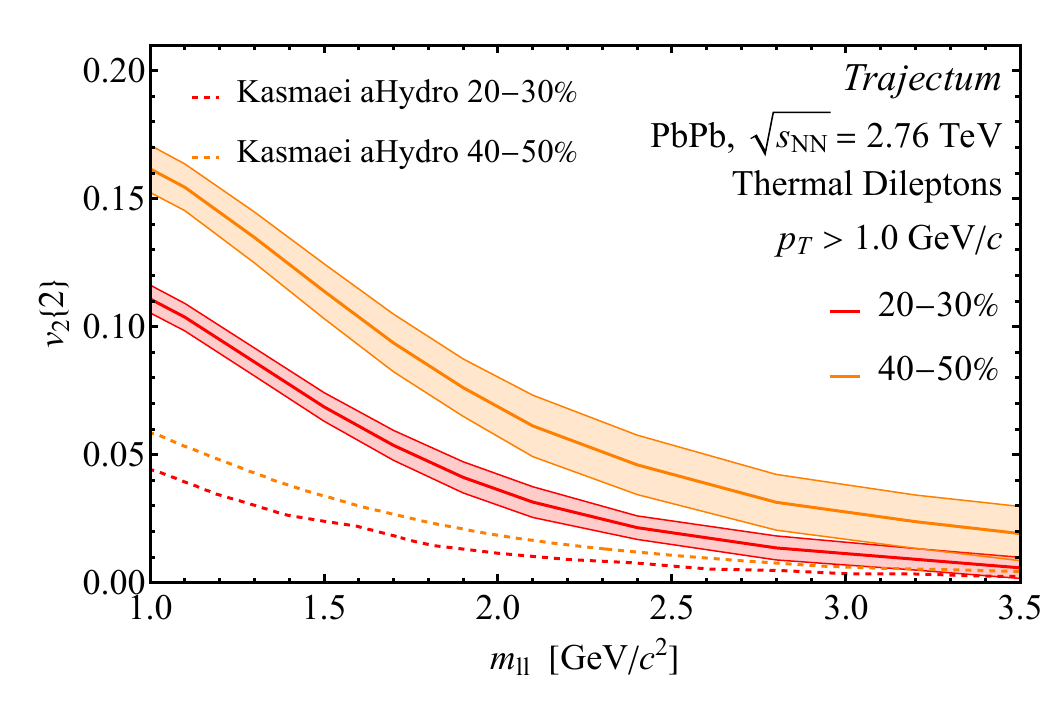}
    \caption{The \textit{Trajectum} results on the elliptic flow ($v_2\{2\}$) of thermal dileptons as a function of $p_T$ (top) and invariant mass spectrum (bottom) compared to \cite{Kasmaei:2018oag}\@, where the kinematic cuts for the results of this work are adjusted such that they match.}
    \label{fig:theorycompdileptonflow}
\end{figure}

In \cite{Linnyk:2015tha}, a different method of modelling relativistic heavy ion collisions from a hydrodynamic approach was used. They calculated the production of direct photons in a relativistic transport approach, describing both the partonic and the hadronic phase: the Parton-Hadron-String Dynamics (PHSD)\@. For the production of direct photons, they covered a wide range of sources: prompt photons from pQCD, radiated photons from quarks and gluons in a strongly interacting QGP, photon radiation by colliding hadronic charges and binary hadron collisions. As can be seen in Fig.~\ref{fig:theorycompphotonspectra}, the direct photon spectra from the PHSD model are comparable to the \emph{Trajectum} results for lower $p_T$\@. However, towards higher $p_T$ the PHSD model predicts significantly higher yields. This is to be expected as for higher momenta the prompt production of photons plays an increasingly important role and we do not take that contribution into account in our results.

The third theoretical model to which we compare our results for thermal photon production is the thermal fireball model from \cite{vanHees:2014ida}\@. It is based on an isotropically expanding cylinder, together with a lattice QCD based equation of state and a sequential freeze-out. The thermal production of photons is based on the full LO results from \cite{Arnold:2001ba} and hadronic many-body interactions. Next to the thermal production they have implemented the prompt production of photons in initial binary collisions as well. In Fig.~\ref{fig:theorycompphotonspectra} we can see that the direct photon production from the fireball model predicts larger yields than \emph{Trajectum} (thermal only) or \cite{Paquet:2015lta}\@.

Fig.~\ref{fig:theorycompphotonflow} compares the thermal photon elliptic flow from this work again to the calculations in \cite{vanHees:2014ida, Linnyk:2015tha, Paquet:2015lta}\@. The inclusion of prompt photons lowers the $v_2$ significantly, especially at larger $p_T$\@. For the thermal component we see that \emph{Trajectum} results agree well with Van Hees et al.~\cite{vanHees:2014ida}, but are lower than the results from Paquet et al.~\cite{Paquet:2015lta}\@. We see that the PHSD model from Linnyk et al.~predicts a significantly lower $v_2$ than \cite{Paquet:2015lta} and a noticeably weaker $p_T$ dependence than either model.

Fig.~\ref{fig:theorycompdileptonspectra} shows the thermal dilepton $p_T$-spectrum and invariant mass spectrum from this work compared to the results from \cite{Kasmaei:2018oag, Garcia-Montero:2024lbl}\@. The results of this work should closely match the results for the thermal contribution found in \cite{Garcia-Montero:2024lbl}\@. In their work they have calculated the invariant mass spectrum on an event-by-event basis using \emph{Trajectum}, using the same MAP parameters as in this work. However, there are some key differences. First of all they have calculated the pre-equilibrium contribution to the direct dilepton mass spectrum, which is not present in our work. The transition from pre-equilibrium to thermal dilepton production takes place at $\tau = 1.0$\,fm$/c$, whereas our thermal production starts at earlier times. In addition they have worked with the LO thermal rates where in our results we have implemented the NLO thermal production rates. Finally in our work we have an additional kinematic cut on the transverse momentum of the pair. As can be seen in the bottom panel of Fig.~\ref{fig:theorycompdileptonspectra}, our calculated QGP contribution is higher compared to the QGP contribution in \cite{Garcia-Montero:2024lbl}, especially for the higher masses. This is to be expected as their thermal production commences at a later time. The combined thermal and pre-equilibrium production lies within the systematic uncertainty of our results.

Fig.~\ref{fig:theorycompdileptonflow} shows the $v_2$ of thermal dileptons as a function of both $p_T$ (top) and invariant mass (bottom), compared to results from \cite{Kasmaei:2018oag}\@. An interesting comparison can be made with the results from \cite{Kasmaei:2018oag}, where the dilepton production was computed in an anisotropic hydrodynamical framework \cite{Alqahtani:2017mhy}\@.  Reference \cite{Kasmaei:2018oag} used a simplified geometric description of the initial state, using a smooth Glauber model, and does not take event-by-event fluctuations into account. We compare our results to the results shown in \cite{Kasmaei:2018oag} for their isotropic dilepton production rates, as we have not incorporated any viscous corrections to the dilepton rates in our simulations. In the comparison we have adapted our kinematic cuts in order to match the kinematic cuts in \cite{Kasmaei:2018oag}, i.e.~$p_T>1.0$\,GeV$/c$ and $m_{ll}>1.0$\,GeV$/c^2$\@. Not only do we find higher yields (see Fig.~\ref{fig:theorycompdileptonspectra}), in addition we find considerably larger elliptic flow values. It is interesting to see these large differences, which could be related to the different ways of calculating the dilepton production rates or could also stem from the differences between viscous hydrodynamics and anisotropic hydrodynamics.

\end{document}